# On propagation of axisymmetric waves in pressurized functionally graded elastomeric hollow cylinders

Bin Wu[1], Yipin Su[1], Dongying Liu[2], Weiqiu Chen[1,3,4,5,*], Chuanzeng Zhang[6,*]

[1] Department of Engineering Mechanics, Zhejiang University, Hangzhou 310027, P.R. China

[2] School of Civil Engineering, Guangzhou University, Guangzhou 510006, P.R. China

[3] State Key Laboratory of Fluid Power and Mechatronic Systems, Zhejiang University, Hangzhou 310027, P.R. China

[4] Key Laboratory of Soft Machines and Smart Devices of Zhejiang Province, Zhejiang University, Hangzhou 310027, P.R. China

[5] Soft Matter Research Center, Zhejiang University, Hangzhou 310027, P.R. China

[6] Department of Civil Engineering, University of Siegen, Siegen D-57068, Germany

**Abstract:**

Soft materials can be designed with a functionally graded (FG) property for specific applications. Such material inhomogeneity can also be found in many soft biological tissues whose functionality is only partly understood to date. In this paper, we analyze the axisymmetric guided wave propagation in a pressurized FG elastomeric hollow cylinder. The cylinder is subjected to a combined action of axial pre-stretch and pressure difference applied to the inner and outer cylindrical surfaces. We consider both torsional waves and longitudinal waves propagating in the FG cylinder made of incompressible isotropic elastomer, which is characterized by the Mooney-Rivlin strain energy function but with the material parameters varying with the radial coordinate in an affine way. The pressure difference generates an inhomogeneous deformation field in the FG cylinder, which dramatically complicates the superimposed wave problem described by the small-on-large theory. A particularly efficient approach is hence employed which combines the state-space formalism for the incremental wave motion with the approximate laminate or multi-layer technique. Dispersion relations for the two types of axisymmetric guided waves are then derived analytically. The accuracy and convergence of the proposed approach is validated numerically. The effects of the pressure difference, material gradient, and axial pre-stretch on both the torsional and the longitudinal wave propagation characteristics are discussed in detail through numerical examples. It is found that the frequency of axisymmetric waves depends nonlinearly on the pressure difference and the material gradient, and an increase in the material gradient enhances the capability of the pressure difference to adjust the wave behavior in the FG cylinder. This work provides a theoretical guidance for characterizing FG soft materials by in-situ ultrasonic nondestructive evaluation and for designing tunable waveguides via material tailoring along with an adjustment of the pre-stretch and pressure difference.

---
[*] Corresponding authors. Tel./Fax: +86-571-87951866; E-mail: chenwq@zju.edu.cn (Weiqiu Chen). Tel.: +49-271-7402173; Fax: +49-271-7404073; E-mail: c.zhang@uni-siegen.de (Chuanzeng Zhang).

| Nomenclature glossary | | | |
|---|---|---|---|
| $B_r, B_0, B_t$ | Undeformed, initial (deformed) and current configurations | $\partial B_r, \partial B_0, \partial B_t$ | Boundaries of $B_r$, $B_0$ and $B_t$ |
| $\mathbf{x} = \boldsymbol{\chi}(\mathbf{X})$, $t$ | Finite motion and time | $\mathrm{d}A, \mathrm{d}a, \mathrm{d}a_t$ | Surface elements in $B_r$, $B_0$ and $B_t$ |
| $\mathbf{X}, \mathbf{x}$ | Position vectors in $B_r$ and $B_t$ | $\mathbf{b}, \mathbf{C}$ | Left and right Cauchy-Green strain tensors |
| $\mathbf{N}, \mathbf{n}, \mathbf{n}_t$ | Outward unit normal vectors in $B_r$, $B_0$ and $B_t$ | $\mathbf{F}, J$ | Deformation gradient tensor and its determinant |
| $\boldsymbol{\sigma}, \mathbf{T}$ | Cauchy and nominal stress tensors | $\rho_r, \rho$ | Mass density in $B_r$ and $B_t$ (or $B_0$) |
| $\mathbf{t}^a, \mathbf{t}^A$ | Applied mechanical traction vectors per unit area of $\partial B_t$ and $\partial B_r$ | $q$ | Lagrange multiplier |
| $W(\mathbf{F})$ | Strain energy function per unit reference volume | $\mathbf{I}, \lambda_i$ ($i=1,2,3$) | Identity tensor and principal stretches |
| $W_i$ | Derivative of $W$ with respect to principal stretch $\lambda_i$ | $\sigma_{ii}(i=1,2,3)$ | Principal Cauchy stresses |
| $\mathbf{u} = \dot{\mathbf{x}}, u_i$ | Incremental displacement vector and its components | $\dot{\mathbf{T}}_0, \dot{T}_{0ij}$ | Push-forward version of Langrangian increment $\dot{\mathbf{T}}$ and its components |
| $\dot{q}$ | Incremental Lagrange multiplier | $\mathbf{H}$ | Incremental displacement gradient tensor with respect to $B_0$ |
| $\boldsymbol{\Gamma}, \Gamma_{\alpha i \beta j}$ | Referential elasticity tensor and its components | $\boldsymbol{\Gamma}_0, \Gamma_{0piqj}$ | Instantaneous elasticity tensor and its components |
| $\dot{\mathbf{t}}^A, \dot{\mathbf{t}}_0^A$ | Lagrangian and Eulerian incremental traction vectors on $\partial B_r$ and $\partial B_0$ | $p_a$ | Applied pressure on the boundary $\partial B_0$ |
| $A, B, L, a, b, l$ | Inner and outer radii, and length of undeformed and deformed FG-EHC | $R, \Theta, Z, r, \theta, z$ | Cylindrical coordinates in undeformed and deformed configurations |
| $\lambda, \lambda_z$ | Circumferential and axial stretches | $\lambda_a, \lambda_b$ | Circumferential stretches of inner and outer surfaces |
| $\eta, \bar{\eta}$ | Ratios of outer radius to inner radius of undeformed and deformed FG-EHC | $H, h$ | Thicknesses of undeformed and deformed FG-EHC |
| $\Lambda, \xi$ | Dimensionless radial coordinates of undeformed and deformed FG-EHC | $W^*(\lambda, \lambda_z)$ | Reduced strain energy density function for axisymmetric deformations |
| $\sigma_{rr}, \sigma_{\theta\theta}, \sigma_{zz}$ | Initial principal stress components in $B_0$ | $p_{in}, p_{ou}, \Delta p$ | Pressures on the inner and outer surfaces of FG-EHC in $B_0$ and pressure difference |
| $N$ | Resultant axial force on each end of deformed FG-EHC | $\mu_1(R), \mu_2(R)$ | Material parameters depending on the radial coordinate $R$ |
| $\mu_{10}, \mu_{20}$ | Material constants in affine variations | $\beta_1, \beta_2$ | Material gradient parameters in affine variations |
| $G$ | Axisymmetric deformation parameter | $\mathbf{Y}, \mathbf{M}$ | Incremental state vector and $6 \times 6$ system matrix |
| | defined as $G = a^2 - \lambda_z^{-1} A^2$ | | |
| $S$ | Integration constant defined in Eq. (25) | $\mathbf{M}_{ij}$ | Four partitioned $3\times3$ sub-matrices of $\mathbf{M}$ |
| $\mathbf{Y}_1, \mathbf{M}_1,$ $\mathbf{V}_1, \bar{\mathbf{M}}_1$ | Incremental state vectors and $2\times2$ system matrices for T waves | $\mathbf{Y}_2, \mathbf{M}_2,$ $\mathbf{V}_2, \bar{\mathbf{M}}_2$ | Incremental state vectors and $4\times4$ system matrices for L waves |
| $f_i (i=1-8)$ | Material parameters appearing in system matrix $\mathbf{M}$ | $U_r, U_\theta, U_z,$ $\Sigma_{0rr}, \Sigma_{0r\theta}, \Sigma_{0rz}$ | Dimensionless modal distribution of incremental fields along radial coordinate |
| $p_0$ | Pressure-like quantity | $n$ | Number of divided sub-layer of FG-EHC |
| $\bar{K}, \bar{\Omega}$ | Dimensionless wave number and circular frequency depending on deformations | $k, K$ | Axial wave number of incremental waves and its dimensionless counterpart |
| $c, v_p$ | Phase velocity of incremental waves and its dimensionless counterpart | $\omega, \Omega$ | Circular frequency of incremental waves and its dimensionless counterpart |
| $\mathbf{S}_k (k=1,2)$ | Global transfer matrices for T and L waves | $\bar{p}_{in}, \bar{p}_{ou}$ | Dimensionless inner and outer pressures |
| $\xi_{j0}, \xi_{j1}, \xi_{jm}$ | Dimensionless radial coordinate at inner/outer/middle surfaces of the $j$th layer | $\mathbf{V}_k(1), \mathbf{V}_k(\bar{\eta})$ | Incremental state vectors at inner and outer surfaces of deformed FG-EHC |
| $\mathbf{D}_k (k=1,2)$ | Coefficient matrices for T and L waves in Eq. (50) | $\mathbf{D}_2^{in}$ | Coefficient matrix for the L wave in Eq. (54) without external pressure |
| $\Delta p^*$ | Dimensionless pressure difference | $\Delta p_c^*$ | Critical pressure difference when FG-EHC collapses |
| $(\Delta p_1^*, \Delta p_2^*)$ | Allowable range of the pressure difference corresponding to $\beta = -0.5$ | $c_T$ | Transverse wave velocity in the undeformed, homogeneous EHC |
| $U_r^*, U_\theta^*, U_z^*$ | Normalized displacement amplitudes of incremental axisymmetric waves | $u, v, w, \dot{Q}$ | Modal components of the incremental displacements and Lagrange multiplier along radial coordinate in Eq. (B.1) |
| $m$ | Circumferential wave number of non-axisymmetric waves (integer) | $\delta_1, \delta_2, \delta_3$ | Terms defined in Eq. (B.4) associated with the Lagrange multiplier |

# 1. Introduction

Owing to their specific mechanical, physical and chemical properties, soft materials such as rubber and rubber-like materials have found various applications, such as in civil and aerospace engineering, automotive industry, seismology, oil prospecting, non-destructive evaluation, signal processing in electronic devices, phononic crystals and metamaterials, underwater engineering, biomechanics and biomedical devices [1-6]. Soft materials often undergo a large deformation, which can be well described by the hyperelasticity theory, and they also operate in a pre-stressed/deformed state. As typical examples, the seismic shock absorbers and the bridge vibration isolators, which protect buildings and highway bridges from earthquake, are rubber components working in a deformed state [1,3]. In biology, a wide range of soft biological tissues have been found to function in a deformed state induced by growth or other biological processes [5]. Recently, soft phononic crystals and metamaterials have been designed to tune and switch acoustic performances upon the applied deformations [2,6]. The presence of pre-stress and pre-deformation may significantly alter the dynamic mechanical characteristics of soft materials.

Consequently, understanding wave propagation behavior in deformed soft materials has not only general theoretical significance but also specific practical importance.

In the context of linear elasticity, wave propagation has been a subject of intensive research interest (see the books by Achenbach [7], Nayfeh [8], and Royer and Dieulesaint [9], to name a few). The investigation of linear incremental waves propagating in deformed soft materials exhibiting both geometric and material nonlinearities should base on the small-on-large theory, which assumes that a time-dependent and small incremental motion is superimposed on a finite initial deformation [1,10]. A large number of studies have been carried out in recent years, and the interested reader is referred to the books of Destrade and Saccomandi [1] and Akbarov [3]. While the former reference provides a unique and multidisciplinary overview on the subject of linear, linearized, and nonlinear waves in pre-stressed soft materials, the latter focuses on the dynamics of pre-strained bi-material systems.

As a common structural element, the particular configuration of cylindrical shells made of soft materials has been used in many engineering applications such as pressure vessels, submarine devices, automobile tires and pipes, tube actuators, phononic crystal and metamaterial systems, and many other structures. These structures are usually subjected to pre-stresses and/or pre-deformations due to environmental pressures or intentionally applied deformations and electric voltages. In addition, many soft biological tissues (such as arterial walls, veins, blood vessels, tendons, etc.) can be modelled as pre-stressed hyperelastic cylindrical structures. Therefore, in order to evaluate the material properties or the pre-stressed state, detect the structural defects and fatigue cracks, realize tunable acoustic devices, and carry out the ultrasound elastography of soft thin-walled biological tissues, much effort has been devoted to the investigation of guided wave propagations in soft hollow cylinders with a pre-deformation within the framework of nonlinear elasticity theory [1,4,5,11-13]. In fact, the small-on-large theory [1,10] has been employed to investigate the effect of a uniform extension or compression on the propagation of torsional, longitudinal, flexural or circumferential waves in hyperelastic solid or hollow cylinders [4,14-22]. These researches well extended the earlier studies by Mindlin and McNiven [23], Pao and Mindlin [24], Onoe et al. [25], and Liu and Qu [26] based on linear elasticity theory. It has been recognized that the wave propagation characteristics in hyperelastic solid or hollow cylinders could be readily changed through the applied pre-stretch. Nonetheless, the pre-stretches/pre-stresses in all the above-mentioned works are assumed to be homogeneous such that the only effect of the pre-stress is to induce material anisotropy. In such cases, exact solutions for various types of the elastic waves have been derived and presented. The first piece of work concerning inhomogeneous pre-deformations is presumably attributed to Engin and Suhubi [27], who investigated the torsional oscillations of an infinite cylindrical elastic tube under large internal and external pressures using the Frobenius method and the variational approach. An application of the pressure difference will result in radially inhomogeneous deformation and stress fields, and hence it is intractable to obtain exact analytical solutions. Utilizing the Liouville-Green method, Shearer et al. [28] re-examined the torsional wave propagation in a pre-stressed hyperelastic hollow cylinder under the combined action of the pressure difference and axial pre-stretch. They pointed out that an unphysical incremental boundary condition had been used in the analysis of Engin and Suhubi [27], resulting in erroneous conclusions that the fundamental mode is deduced to be dispersive and does not pass through the origin. In fact, Shearer et al. [28] demonstrated that the fundamental mode of the torsional wave should be always nondispersive and pass through the origin. Interestingly, the inhomogeneous deformations and stresses generated by the combination of the pressure difference and axial pre-stretch have been used to realize hyperelastic cloaking and tunable elastodynamic band gaps of anti-plane elastic waves [11,12]. More recently, Shmuel and deBotton [29], Shmuel [30], and Wu et al. [4] considered respectively the axisymmetric, torsional, and circumferential waves propagating in soft electroactive tubes subjected to a radially applied electric field as well as an axial mechanical force. The application of an electric voltage difference to the inner and outer surfaces of the soft electroactive tube will also result in radially inhomogeneous deformation fields, similar to the case of a radial pressure difference.

In the middle of 1980s, a new type of materials called functionally graded materials (FGMs), which exhibit inhomogeneous material properties varying continuously along one or more directions, was first proposed and prepared by a group of Japanese material scientists. FGMs have been designed for various purposes of reducing the residual and thermal stresses, increasing the bonding strength, eliminating the sharp stress discontinuity, and improving the efficiency and life of acoustic wave devices [31,32]. Because of the excellent performance and superiority over many other conventional materials, FGMs have gained widespread applications not only in aerospace engineering but also in biology [33,34]. In the linear elasticity context, extensive researches have been conducted on the thermomechanical responses and fracture behavior of FGMs over the past years [35,36]. For the purpose of the material characterization via ultrasonic nondestructive evaluation, it is very important to investigate the wave propagation in FGMs. However, owing to the spatially varying material properties, the resulting displacement equations of motion in general are a system of coupled partial differential equations with variable coefficients [31,37], which are difficult to solve analytically or numerically, even in the elastic regime. In order to alleviate this difficulty, many works have been carried out on developing numerical techniques to solve wave propagation and transient response problems in FG structures. For example, using the linearly inhomogeneous elements and quadratic layer elements, respectively, Liu et al. [31] and Han et al. [38] investigated the stress wave propagation in FGMs and pointed out the possibility of wave-based material characterization. Furthermore, Han et al. [39,40] studied the wave propagation characteristics and transient responses in FG cylinders employing a hybrid numerical method. To find an appropriate method for studying the wave dispersion behavior in FG plates, Chen et al. [37] comprehensively presented a comparison of three different methods, i.e., the conventional displacement method, the state-space method (SSM), and the reverberation-ray matrix method (MRRM), and demonstrated that the MRRM is particularly suitable for the calculation at large wave numbers due to its numerical stability while the SSM is suitable for the calculation at small wave numbers. Zhu et al. [41] further proposed a recursive formula for the MRRM to analyze free wave propagation in an FG elastic plate, with much higher computational efficiency than the traditional MRRM developed for studying transient waves in frames [42].

Although the concept of FGMs is well-established and numerous works on FGMs have been conducted, its application and generalization to soft materials with highly elastic behavior is relatively few. In fact, spatial variations or inhomogeneities in chemical and mechanical properties can be readily introduced in rubber-like materials either from the non-uniform transport variables like temperature or from the adverse action of thermo-oxidation [43]. Besides, FG rubber-like materials were first prepared by Ikeda et al. [44] using a construction-based (layering) method in the laboratory. Furthermore, a lot of natural soft biological tissues (such as blood vessels, articular cartilages, arterial walls, tendons, etc.) have an intricate architecture and typically exhibit FG material properties [45]. By using a hierarchical reinforcement approach, heterogeneous composites exhibiting extreme soft-to-hard transition regions and still being reversibly stretchable up to 350% were designed and produced for applications ranging from flexible electronics to regenerative medicine [46]. A combustion-powered robot, whose body transitions from a rigid core to a soft exterior with a stiffness gradient spanning three orders of magnitude in modulus, was manufactured by Bartlett et al. [47] by means of a multi-material 3D printing. In theoretical aspects, nonlinear deformations including axial extension or compression, inflation, torsion, and shearing of incompressible or compressible FG soft hollow cylinders have gained a considerable research attention [48-52]. By properly tailoring the material properties in one or more directions, the working performance of soft FGMs can be optimized. Batra and Bahrami [51] studied the axisymmetric radial deformations of an FG Mooney-Rivlin hollow cylinder with the two material parameters varying continuously through the thickness described by either a power law or an affine relation. Saravanan and Rajagopal [50] have showed that the homogenization approach obtained using the equivalent energy principle may lead to serious errors in the estimation of the stresses or tractions in an inhomogeneous soft material since the stresses depend on the derivatives of the stored energy with respect to the deformation gradient, and hence most homogenization procedures cannot develop a reasonable failure criterion for inhomogeneous soft materials. Recently, a systematic instability analysis was carried out by Chen et al. [32] for an FG elastomeric hollow cylinder (FG-EHC) subjected to both end thrust and internal/external pressure. It should be emphasized that, the aforementioned works, primarily in the purely static context, demonstrate a particularly significant role that the material gradient plays in the mechanical responses to external stimuli and the performance of soft materials and structures. To the authors' best knowledge, wave propagation in pre-stressed FG elastomeric materials and structures has not been studied yet in the framework of nonlinear elasticity.

This paper will examine the axisymmetric guided wave propagation in a pressurized FG-EHC, in order to provide a theoretical guidance for characterization of material properties, pre-stress states, and structural defects or cracks by in-situ ultrasonic nondestructive evaluation as well as for design of tunable waveguides via material tailoring along with pre-stretch and pressure adjusting. The paper is organized as follows. Based on the work of Ogden [10], Section 2 first gives a brief overview on the general nonlinear elasticity theory and the linear incremental theory for hyperelastic materials. For an arbitrary strain energy function, the axisymmetric deformation response of an FG-EHC to the combined action of the pre-stretch and pressure difference is analyzed in Section 3 following Batra [52] and Chen et al. [32]. The analysis is then specialized to the Mooney-Rivlin model with affine variations of the material parameters in the radial direction. The state-space formalism for the incremental fields in cylindrical coordinates is formulated in Section 4, which is valid for any strain energy function. An appropriate approach, which combines the approximate laminate or multi-layer technique with the state-space formalism, is utilized in Section 5 to obtain the dispersion relations for both incremental torsional and longitudinal waves (hereafter abbreviated as the T waves and L waves, respectively) propagating in the pressurized FG-EHC with radially inhomogeneous pre-deformations. Numerical examples are finally presented to illustrate the nonlinear static response of the pressurized FG-EHC in Section 6.1 and the wave propagation analysis in Section 6.2 including the analysis of the validation of the proposed solution technique in Section 6.2.1 as well as the effects of the material gradient, pressure difference, and axial pre-stretch on the T and L waves in Sections 6.2.2 and 6.2.3, respectively. Section 7 is devoted to a concluding summary. Some relevant mathematical expressions or derivations are provided in Appendices A and B.

## 2. Theoretical background

In this section, the nonlinear elasticity theory of a soft deformable continuum as well as the linearized theory for the incremental deformation superimposed on a finitely deformed configuration will be briefly reviewed. The interested readers are referred to the monograph of Ogden [10] as well as the references cited therein for more detailed discussions about the basic ideas and formulations.

### 2.1. Nonlinear elasticity theory

In the undeformed stress-free reference configuration, the soft deformable continuous body occupies a region $B_r$ with boundary $\partial B_r$ and outward unit normal $\mathbf{N}$. An arbitrary material point in this state labelled as $X$ is identified by its position vector $\mathbf{X}$. An application of the external stimuli deforms the body such that the material point $X$ at time $t$ occupies a new position $\mathbf{x} = \boldsymbol{\chi}(\mathbf{X}, t)$ in the deformed or current configuration $B_t$ with boundary $\partial B_t$ and outward unit normal $\mathbf{n}_t$. Here, $\boldsymbol{\chi}$ maps $\mathbf{X}$ to $\mathbf{x}$ and is an invertible vector function. The deformation gradient tensor relative to $B_r$ is defined as $\mathbf{F} = \partial \mathbf{x} / \partial \mathbf{X} = \text{Grad}\,\boldsymbol{\chi}$, where Grad is the gradient operator with respect to $B_r$. The local measure of the volume change is $J = \det \mathbf{F} = 1$ for an incompressible material. The left and right Cauchy-Green strain tensors $\mathbf{b} = \mathbf{F}\mathbf{F}^{\mathrm{T}}$ and $\mathbf{C} = \mathbf{F}^{\mathrm{T}}\mathbf{F}$ will be used as the deformation measures.

In Lagrangian and Eulerian forms, the equations of motion for the continuum in the absence of body forces are respectively

$$\text{Div}\mathbf{T} = \rho_r \mathbf{x}_{,tt} \quad \text{or} \quad \text{div}\boldsymbol{\sigma} = \rho \mathbf{x}_{,tt} \tag{1}$$

where the subscript $t$ following a comma denotes the material time derivative, Div and div are the divergence operators with respect to $B_r$ and $B_t$, respectively, and the nominal stress tensor $\mathbf{T}$ is related to the Cauchy stress tensor $\boldsymbol{\sigma}$ by $\mathbf{T} = \mathbf{F}^{-1}\boldsymbol{\sigma}$ for incompressible materials. The conservation of angular momentum leads to the symmetry of $\boldsymbol{\sigma}$. In Eq. (1), $\rho_r$ and $\rho$ are the mass density of the material in $B_r$ and $B_t$, respectively, which are equal during the motion because of the material incompressibility.

In terms of the nominal and Cauchy stress tensors, the mechanical boundary conditions may be written in Lagrangian and Eulerian forms respectively as

$$\begin{aligned} \mathbf{T}^{\mathrm{T}}\mathbf{N} &= \mathbf{t}^A, \quad \text{on} \quad \partial B_r \\ \boldsymbol{\sigma}\mathbf{n}_t &= \mathbf{t}^a, \quad \text{on} \quad \partial B_t \end{aligned} \tag{2}$$

where the superscript T signifies the transpose, and $\mathbf{t}^A$ and $\mathbf{t}^a$, which are related through $\mathbf{t}^A \mathrm{d}A = \mathbf{t}^a \mathrm{d}a_t$, denote the applied mechanical traction vectors per unit area of $\partial B_r$ and $\partial B_t$, respectively. The well-known Nanson's formula $\mathbf{n}_t \mathrm{d}a_t = \mathbf{F}^{-\mathrm{T}}\mathbf{N}\mathrm{d}A$ for incompressible materials is used to connect the surface elements $\mathrm{d}A$ and $\mathrm{d}a_t$.

The hyperelasticity theory describes the nonlinear elastic responses of a material with a strain energy function $W(\mathbf{F})$ per unit reference volume. Following Ogden [10], the nonlinear constitutive relations for incompressible elastic materials in terms of the nominal and Cauchy stress tensors are given respectively by

$$\mathbf{T} = \frac{\partial W}{\partial \mathbf{F}} - q\mathbf{F}^{-1} \quad \text{or} \quad \boldsymbol{\sigma} = \mathbf{F}\frac{\partial W}{\partial \mathbf{F}} - q\mathbf{I} \tag{3}$$

where $q$ is a Lagrange multiplier associated with the incompressibility constraint and $\mathbf{I}$ is the identity tensor. For an isotropic elastic material, the strain energy function $W$ depends symmetrically on the principal stretches $\lambda_i$ $(i = 1, 2, 3)$ such that $W = W(\lambda_1, \lambda_2, \lambda_3)$ with $\lambda_1 \lambda_2 \lambda_3 = 1$ due to the material incompressibility. Therefore, referring to the principal axes of $\boldsymbol{\sigma}$, the corresponding principal Cauchy stresses $\sigma_{ii}$ $(i = 1, 2, 3)$ are given by

$$\sigma_{ii} = \lambda_i W_i - q, \quad i = 1, 2, 3 \text{ (no summation)} \tag{4}$$

where the notation $W_i = \partial W / \partial \lambda_i$ is introduced.

## 2.2. The linearized incremental theory

Now suppose that a time-dependent infinitesimal incremental deformation $\dot{\mathbf{x}}(\mathbf{X}, t)$ is superimposed on an underlying static configuration $B_0$ (with the boundary $\partial B_0$, the surface element $\mathrm{d}a$, and the outward unit normal vector $\mathbf{n}$) which carries a finite deformation $\mathbf{x} = \boldsymbol{\chi}(\mathbf{X})$. The increments are signified by superposed dots. Note here that it is unnecessary to distinguish between the underlying initial deformed configuration $B_0$ and the current configuration $B_t$ since the incremental deformation is assumed to be infinitesimal. According to Ogden [10] and Wu et al. [53], the governing equations for the linearized incremental theory can be established by using the perturbation method and expressed in Eulerian and Lagrangian forms, respectively, when the reference configuration is updated from the undeformed configuration $B_r$ to the deformed configuration $B_0$. This linearized incremental theory is approximate in nature, however, it is sufficiently accurate just like the linear elasticity theory for an infinitesimal deformation of elastic solids [4,10]. In the following, we will only review the linearized incremental theory in Eulerian form which will be utilized later to investigate the axisymmetric wave propagation in the pressurized FG-EHC.

The incremental equations of motion are written as

$$\operatorname{div} \dot{\mathbf{T}}_0 = \rho \mathbf{u}_{,tt} \tag{5}$$

where $\dot{\mathbf{T}}_0 = \mathbf{F} \dot{\mathbf{T}}$ is the push-forward version of the corresponding Lagrangian increment $\dot{\mathbf{T}}$ for incompressible materials and the incremental displacement vector $\mathbf{u}(\mathbf{x})$ is defined through $\mathbf{u}(\mathbf{x}) = \mathbf{u}(\boldsymbol{\chi}(\mathbf{X})) = \dot{\mathbf{x}}(\mathbf{X})$. The subscript 0 is used to identify the resulting push-forward variables. As the first-order approximation, the incremental stress tensor $\dot{\mathbf{T}}_0$ is calculated by the following linearized incremental constitutive laws for incompressible elastic materials

$$\dot{\mathbf{T}}_0 = \boldsymbol{\Gamma}_0 \mathbf{H} + q\mathbf{H} - \dot{q}\mathbf{I} \tag{6}$$

where $\dot{q}$ is the incremental Lagrange multiplier and $\mathbf{H} = \operatorname{grad} \mathbf{u}$ denotes the displacement gradient tensor with grad being the gradient operator with respect to $B_0$. The fourth-order instantaneous elasticity tensor $\mathbf{\Gamma}_0$ is given in component notation by

$$\Gamma_{0piqj} = F_{p\alpha} F_{q\beta} \Gamma_{\alpha i\beta j} = \Gamma_{0qjpi} \tag{7}$$

where $\mathbf{\Gamma}$ denotes the referential elasticity tensor associated with $W(\mathbf{F})$, with its components given by $\Gamma_{\alpha i\beta j} = \partial^2 W / (\partial F_{i\alpha} \partial F_{j\beta})$. Here, Greek indices are associated with the reference configuration $B_r$ while Roman indices with $B_0$. By referring to the principal axes of $\boldsymbol{\sigma}$, the non-zero components of $\mathbf{\Gamma}_0$ for incompressible isotropic elastic materials can be written in terms of the strain energy function $W$ and the principal stretches $\lambda_i$ as [10]

$$\begin{aligned}
&\Gamma_{0iiii} = \Gamma_{0jiii} = \lambda_i \lambda_j W_{ij}, \\
&\Gamma_{0ijij} = \begin{cases}
\dfrac{\lambda_i W_i - \lambda_j W_j}{\lambda_i^2 - \lambda_j^2} \lambda_i^2, & (i \neq j, \ \lambda_i \neq \lambda_j) \\
\dfrac{1}{2}\left(\Gamma_{0iiii} - \Gamma_{0iijj} + \lambda_i W_i\right), & (i \neq j, \ \lambda_i = \lambda_j)
\end{cases} \\
&\Gamma_{0ijji} = \Gamma_{0jiij} = \Gamma_{0ijij} - \lambda_i W_i, \ (i \neq j)
\end{aligned} \tag{8}$$

where $W_{ij} = \partial^2 W / \partial \lambda_i \partial \lambda_j$. In addition, the incremental incompressibility condition is given by

$$\mathrm{div}\,\mathbf{u} = 0 \tag{9}$$

The Eulerian incremental form of the traction boundary conditions, which are to be satisfied on $\partial B_0$, can be written as

$$\dot{\mathbf{T}}_0^{\mathrm{T}} \mathbf{n} = \dot{\mathbf{t}}_0^A \tag{10}$$

where we have the relation $\dot{\mathbf{t}}_0^A \mathrm{d}a = \dot{\mathbf{t}}^A \mathrm{d}A$ based on the Nanson's formula, with $\dot{\mathbf{t}}^A$ being the Lagrangian incremental traction vector per unit area of $\partial B_r$. When the boundary $\partial B_0$ is subjected to an applied pressure $p_a$, the incremental boundary conditions (10) can be written as [32,54]

$$\dot{\mathbf{T}}_0^{\mathrm{T}} \mathbf{n} = p_a \mathbf{H}^{\mathrm{T}} \mathbf{n} - \dot{p}_a \mathbf{n} \tag{11}$$

## 3. Axisymmetric deformation of a pressurized FG-EHC

For an arbitrary strain energy function, the problem of the axisymmetric deformation of a pressurized FG-EHC subjected to an axial tension or compression was investigated by Batra [52] and Chen et al. [32]. Here we provide a brief summary of the main results needed for our analysis. When the FG-EHC is in the undeformed configuration (Fig. 1(a)), its inner and outer radii are $A$ and $B$, and its length is $L$. Then both internal and external pressures as well as an axial tension or compression are applied to the FG-EHC (Fig. 1(b)) and the corresponding geometrical parameters will become $a$, $b$ and $l$, respectively. An axisymmetric deformation implies that the shape of the circular cylinder is maintained during the loading process. Therefore, two cylindrical coordinate systems $(R, \Theta, Z)$ and $(r, \theta, z)$ with unit base vectors $\mathbf{e}_1, \mathbf{e}_2, \mathbf{e}_3$ corresponding to the $r, \theta, z$-coordinates are adopted to describe the motion, as shown in Fig. 1.

Under the assumption that the material is incompressible, the axisymmetric deformation is described by [4,32]

$$R^2 - A^2 = \lambda_z (r^2 - a^2), \quad \theta = \Theta, \quad z = \lambda_z Z \tag{12}$$

where $\lambda_z \equiv \lambda_3 = l / L$ is the axial principal stretch, which is independent of $r$. Therefore, from the incompressibility condition, the radial and circumferential stretches are $\lambda_1 = \mathrm{d}r / \mathrm{d}R = \lambda^{-1} \lambda_z^{-1}$ and $\lambda_2 = r / R \equiv \lambda$, respectively. For convenience, we introduce the following notational convention

$$\begin{aligned} \lambda_a = a / A, \quad \lambda_b = b / B, \quad \eta = B / A, \quad \overline{\eta} = b / a, \\ \Lambda = R / A, \quad \xi = r / a, \quad H = B - A, \quad h = b - a \end{aligned} \tag{13}$$

Substituting Eq. (13) into Eq. (12), we obtain

$$\lambda_a^2 \lambda_z - 1 = \Lambda^2 \left( \lambda^2 \lambda_z - 1 \right) = \eta^2 \left( \lambda_b^2 \lambda_z - 1 \right) \tag{14}$$

Note that under the incompressibility constraint it is convenient to define a reduced strain energy function $W^*$ as $W^*(\lambda, \lambda_z) = W(\lambda^{-1} \lambda_z^{-1}, \lambda, \lambda_z)$, and thus the principal stress differences can be written from Eq. (4) as

$$\sigma_{\theta\theta} - \sigma_{rr} = \lambda W_\lambda^*, \quad \sigma_{zz} - \sigma_{rr} = \lambda_z W_{\lambda_z}^* \tag{15}$$

where $W_\lambda^* = \partial W^* / \partial \lambda$ and $W_{\lambda_z}^* = \partial W^* / \partial \lambda_z$.

For the axisymmetric deformation considered here, all physical quantities depend only on $r$. Therefore, in view of Eq (15)$_1$, the only equilibrium equation $\mathrm{div}\boldsymbol{\sigma} = \mathbf{0}$ not satisfied trivially is

$$\mathrm{d}\sigma_{rr} / \mathrm{d}r = \lambda W_\lambda^* / r \tag{16}$$

The corresponding boundary conditions on the inner and outer cylindrical surfaces can be written as

$$\sigma_{rr}|_{r=a} = -p_{in}, \quad \sigma_{rr}|_{r=b} = -p_{ou} \tag{17}$$

Consequently, the integration of Eq. (16) from $a$ to $b$ leads to

$$p_{ou} - p_{in} = -\int_a^b \lambda W_\lambda^* \frac{\mathrm{d}r}{r} = \int_{\lambda_a}^{\lambda_b} \frac{W_\lambda^*}{\lambda^2 \lambda_z - 1} \mathrm{d}\lambda \tag{18}$$

where the identity $\mathrm{d}r / r = -\mathrm{d}\lambda / [\lambda(\lambda^2 \lambda_z - 1)]$ has been used based on the definition $\lambda = r / R$.
Since $\lambda_b$ can be expressed in terms of $\lambda_a$ and $\lambda_z$ by Eq. (14), Eq. (18) establishes a general
nonlinear relation between the pressure difference (or the net pressure) $\Delta p = p_{ou} - p_{in}$ and the
inner radius $a$ (or $\lambda_a$) once the geometric parameter $\eta$ and the axial stretch $\lambda_z$ are prescribed.

The radial normal stress can be found by integrating Eq. (16) from $a$ to $r$ as

$$\sigma_{rr}(r) = -p_{in} + \int_a^r \lambda W_\lambda^* \frac{\mathrm{d}r}{r} = -p_{in} - \int_{\lambda_a}^{\lambda} \frac{W_\lambda^*}{\lambda^2 \lambda_z - 1} \mathrm{d}\lambda \tag{19}$$

In terms of Eq. (15)$_2$, the integration of the axial normal stress $\sigma_{zz}$ over the cross-section of the
deformed FG-EHC yields the resultant axial force

$$N = 2\pi \int_a^b \sigma_{zz}(r) r \mathrm{d}r = 2\pi \int_a^b \left( \sigma_{rr} + \lambda_z W_{\lambda_z}^* \right) r \mathrm{d}r \tag{20}$$

It should be emphasized here that the theoretical formulation described above is completely
general for an incompressible isotropic FG-EHC without any specialization on the form of the
reduced strain energy function $W^*$. The integrations (18)-(20) can be performed analytically or
numerically when the form of $W^*$ is given.

In this paper, the FG-EHC is assumed to be characterized by the incompressible Mooney-
Rivlin model [32,51] with the following strain energy function:

$$W(\lambda_1, \lambda_2, \lambda_3) = \mu_1(R)(\lambda_1^2 + \lambda_2^2 + \lambda_3^2 - 3) / 2 - \mu_2(R)(\lambda_1^{-2} + \lambda_2^{-2} + \lambda_3^{-2} - 3) / 2 \tag{21}$$

which can be written in the reduced form as

$$W^*(\lambda, \lambda_z) = \mu_1(R)(\lambda^{-2}\lambda_z^{-2} + \lambda^2 + \lambda_z^2 - 3) / 2 - \mu_2(R)(\lambda^2 \lambda_z^2 + \lambda^{-2} + \lambda_z^{-2} - 3) / 2 \tag{22}$$

where $\mu_1(R)$ and $\mu_2(R)$ are the material parameters that depend on the radial coordinate $R$. For
an infinitesimal deformation, $\mu_1(R) - \mu_2(R)$ corresponds to the shear modulus $\mu(R)$ of the
undeformed state. To be more specific, we only consider the following affine variations of
$\mu_1(R)$ and $\mu_2(R)$:

$$\mu_1(R) = \mu_{10}\left(1 + \beta_1 R / A\right) = \mu_{10}\left(1 + \beta_1 \Lambda\right), \quad \mu_2(R) = \mu_{20}\left(1 + \beta_2 R / A\right) = \mu_{20}\left(1 + \beta_2 \Lambda\right) \qquad (23)$$

where $\mu_{10}$ and $\mu_{20}$ are the material constants with unit N/m$^2$, and $\beta_1$ and $\beta_2$ are two dimensionless parameters characterizing the functionally graded property of the soft material. For the functionally graded Mooney-Rivlin model (21)-(23), Batra and Bahrami [51] and Chen et al. [32] have analytically determined the nonlinear response for the axisymmetric deformation and derived the explicit expressions for the radially inhomogeneous physical variables in the pressurized FG-EHC. We cite their results below, but use our notation.

First, the radially inhomogeneous physical variables are given by

$$\begin{aligned}
\sigma_{rr} &= -p_{in} - \frac{\mu_{10}}{\lambda_z}\left[\ln\lambda + \frac{G(1+\beta_1\Lambda)}{2r^2} - \frac{3\beta_1\sqrt{G\lambda_z}}{2A}\tan^{-1}\left(\frac{R}{\sqrt{G\lambda_z}}\right)\right] \\
&\quad + \mu_{20}\lambda_z\left[\ln\lambda + \frac{G(1+\beta_2\Lambda)}{2r^2} - \frac{3\beta_2\sqrt{G\lambda_z}}{2A}\tan^{-1}\left(\frac{R}{\sqrt{G\lambda_z}}\right)\right] + S, \\
\sigma_{\theta\theta} &= \sigma_{rr} + \mu_{10}\left(1+\beta_1\Lambda\right)\left(\lambda^2 - \lambda_z^{-2}\lambda^{-2}\right) + \mu_{20}\left(1+\beta_2\Lambda\right)\left(\lambda^{-2} - \lambda_z^2\lambda^2\right), \\
\sigma_{zz} &= \sigma_{rr} + \mu_{10}\left(1+\beta_1\Lambda\right)\left(\lambda_z^2 - \lambda_z^{-2}\lambda^{-2}\right) + \mu_{20}\left(1+\beta_2\Lambda\right)\left(\lambda_z^{-2} - \lambda_z^2\lambda^2\right), \\
q &= \mu_{10}\left(1+\beta_1\Lambda\right)\lambda_z^{-2}\lambda^{-2} + \mu_{20}\left(1+\beta_2\Lambda\right)\lambda_z^2\lambda^2 - \sigma_{rr}
\end{aligned} \qquad (24)$$

where $r = \sqrt{G + \lambda_z^{-1}R^2}$, $G = a^2 - \lambda_z^{-1}A^2$, and

$$\begin{aligned}
S &= \frac{\mu_{10}}{\lambda_z}\left[\ln\lambda_a + \frac{G(1+\beta_1)}{2a^2} - \frac{3\beta_1\sqrt{G\lambda_z}}{2A}\tan^{-1}\left(\frac{A}{\sqrt{G\lambda_z}}\right)\right] \\
&\quad - \mu_{20}\lambda_z\left[\ln\lambda_a + \frac{G(1+\beta_2)}{2a^2} - \frac{3\beta_2\sqrt{G\lambda_z}}{2A}\tan^{-1}\left(\frac{A}{\sqrt{G\lambda_z}}\right)\right]
\end{aligned} \qquad (25)$$

Furthermore, the equations governing the nonlinear response of the pressurized FG-EHC can be written as

$$\begin{aligned}
\Delta p = p_{ou} - p_{in} &= \frac{\mu_{10}}{\lambda_z}\left[\ln\lambda_b + \frac{G(1+\beta_1\eta)}{2b^2} - \frac{3\beta_1\sqrt{G\lambda_z}}{2A}\tan^{-1}\left(\frac{B}{\sqrt{G\lambda_z}}\right)\right] \\
&\quad - \mu_{20}\lambda_z\left[\ln\lambda_b + \frac{G(1+\beta_2\eta)}{2b^2} - \frac{3\beta_2\sqrt{G\lambda_z}}{2A}\tan^{-1}\left(\frac{B}{\sqrt{G\lambda_z}}\right)\right] - S
\end{aligned} \qquad (26)$$

from which the inner radius $a$ (or $\lambda_a$) can be determined in terms of the pressure difference $\Delta p$ when the geometric parameter $\eta$ and the axial stretch $\lambda_z$ are given. The resultant axial force is

$$N = N_R(R)\big|_B^A \equiv N_R(A) - N_R(B),$$

$$N_R(R) = \frac{2\pi\mu_{10}}{\lambda_z^2}\left[\frac{1}{2}R^2\ln\lambda + \frac{\beta_1 G\lambda_z}{4A}R - \frac{\beta_1\lambda_z}{2A}\sqrt{G\lambda_z}\left(\frac{3}{2}r^2 - G\right)\tan^{-1}\left(\frac{R}{\sqrt{G\lambda_z}}\right)\right]$$

$$-2\pi\mu_{20}\left[\left(\frac{1}{2}R^2 + G\lambda_z\right)\ln\lambda + \frac{\beta_2 G\lambda_z}{4A}R - \frac{\beta_2\lambda_z}{2A}\sqrt{G\lambda_z}\left(G + \frac{3}{2}r^2\right)\tan^{-1}\left(\frac{R}{\sqrt{G\lambda_z}}\right)\right] \qquad (27)$$

$$-2\pi\left(1 - \lambda_z^{-3}\right)\left[\mu_{10}\lambda_z\left(\frac{R^2}{2} + \frac{\beta_1 R^3}{3A}\right) - \mu_{20}\left(\frac{R^2}{2} + \frac{\beta_2 R^3}{3A}\right)\right] - \frac{\pi\left(S - p_{in}\right)R^2}{\lambda_z}$$

Eqs. (24)-(27) can be readily obtained by substituting Eq. (21) or Eq. (22) into the general expressions (4) and (15)-(20) and then integrating the resulting equations.

If the material parameters $\mu_1(R)$ and $\mu_2(R)$ in the incompressible Mooney-Rivlin model (21) are assumed to be constant, i.e., $\beta_1 = \beta_2 = 0$, the radial normal stress can be written according to Eqs. (24)$_1$ and (25) as

$$\sigma_{rr} = -p_{in} - \left(\frac{\mu_{10}}{\lambda_z} - \mu_{20}\lambda_z\right)\left[\ln\frac{\lambda}{\lambda_a} + \frac{G}{2}\left(\frac{1}{r^2} - \frac{1}{a^2}\right)\right] \qquad (28)$$

Accordingly, Eq. (26) becomes

$$p_{ou} - p_{in} = \left(\frac{\mu_{10}}{\lambda_z} - \mu_{20}\lambda_z\right)\left[\ln\frac{\lambda_b}{\lambda_a} + \frac{\lambda_z G}{2}\frac{A^2 - B^2}{\left(\lambda_z G + A^2\right)\left(\lambda_z G + B^2\right)}\right] \qquad (29)$$

which is equivalent to the result obtained by Shearer et al. [28] but expressed in different notation. We note that the expression in Shearer et al. [28] contains a typographical error, which has been corrected in Eq. (29).

It can be seen from Eqs. (24) and (28) that the application of an internal or external pressure makes the initial physical variables radially inhomogeneous, even when the material is homogenous. However, as pointed out by Chen et al. [32], when the internal and external pressures both vanish, a uniform deformation state always exists in FG elastomeric hollow or solid cylinders described by an arbitrary radially inhomogeneous Mooney-Rivlin model subjected to an axial tension or compression. Specifically, both the radial and the circumferential stretches are constant and related to the axial stretch by $\lambda_1 = \lambda_2 = \lambda_z^{-1/2}$. In addition, both the radial and the circumferential stresses vanish, and the only nonzero component of the Cauchy stress tensor is the axial stress $\sigma_{zz}$ given by

$$\sigma_{zz} = \mu_1(R)\left(\lambda_z^2 - \lambda_z^{-1}\right) + \mu_2(R)\left(\lambda_z^{-2} - \lambda_z\right) \tag{30}$$

Although the deformation state is uniform, the axial stress $\sigma_{zz}$ varies along the radial direction. The axial stress will also become uniform along the radial direction if the material is homogeneous.

## 4. State-space formalism for incremental fields

Without the need to specify the form of the strain energy function, the incremental governing equations obtained in Section 2.2 can be recast into their equivalent forms in the cylindrical coordinates $(r, \theta, z)$ in order to describe the time-dependent incremental motion superimposed on the previously determined, radially inhomogeneous axisymmetric deformation in the pressurized FG-EHC.

In the cylindrical coordinates, the incremental displacement gradient tensor becomes

$$\mathbf{H} = \begin{bmatrix} \dfrac{\partial u_r}{\partial r} & \dfrac{1}{r}\left(\dfrac{\partial u_r}{\partial \theta} - u_\theta\right) & \dfrac{\partial u_r}{\partial z} \\[2mm] \dfrac{\partial u_\theta}{\partial r} & \dfrac{1}{r}\left(\dfrac{\partial u_\theta}{\partial \theta} + u_r\right) & \dfrac{\partial u_\theta}{\partial z} \\[2mm] \dfrac{\partial u_z}{\partial r} & \dfrac{1}{r}\dfrac{\partial u_z}{\partial \theta} & \dfrac{\partial u_z}{\partial z} \end{bmatrix} \tag{31}$$

along with the incremental incompressibility constraint

$$\frac{\partial u_r}{\partial r} + \frac{1}{r}\left(\frac{\partial u_\theta}{\partial \theta} + u_r\right) + \frac{\partial u_z}{\partial z} = 0 \tag{32}$$

The corresponding incremental equations of motion (5) can be written as

$$\frac{\partial \dot{T}_{0rr}}{\partial r} + \frac{1}{r}\frac{\partial \dot{T}_{0\theta r}}{\partial \theta} + \frac{\dot{T}_{0rr} - \dot{T}_{0\theta\theta}}{r} + \frac{\partial \dot{T}_{0zr}}{\partial z} = \rho\frac{\partial^2 u_r}{\partial t^2}$$

$$\frac{\partial \dot{T}_{0r\theta}}{\partial r} + \frac{1}{r}\frac{\partial \dot{T}_{0\theta\theta}}{\partial \theta} + \frac{\dot{T}_{0\theta r} + \dot{T}_{0r\theta}}{r} + \frac{\partial \dot{T}_{0z\theta}}{\partial z} = \rho\frac{\partial^2 u_\theta}{\partial t^2} \tag{33}$$

$$\frac{\partial \dot{T}_{0rz}}{\partial r} + \frac{1}{r}\frac{\partial \dot{T}_{0\theta z}}{\partial \theta} + \frac{\partial \dot{T}_{0zz}}{\partial z} + \frac{\dot{T}_{0rz}}{r} = \rho\frac{\partial^2 u_z}{\partial t^2}$$

Using Eqs. (8) and (31), the incremental constitutive equations (6) are reduced to

$$\dot{T}_{0rr} = (\Gamma_{01111} + q)\frac{\partial u_r}{\partial r} + \Gamma_{01122}\frac{1}{r}\left(\frac{\partial u_\theta}{\partial \theta} + u_r\right) + \Gamma_{01133}\frac{\partial u_z}{\partial z} - \dot{q},$$

$$\dot{T}_{0\theta\theta} = \Gamma_{01122}\frac{\partial u_r}{\partial r} + (\Gamma_{02222} + q)\frac{1}{r}\left(\frac{\partial u_\theta}{\partial \theta} + u_r\right) + \Gamma_{02233}\frac{\partial u_z}{\partial z} - \dot{q},$$

$$\dot{T}_{0zz} = \Gamma_{01133}\frac{\partial u_r}{\partial r} + \Gamma_{02233}\frac{1}{r}\left(\frac{\partial u_\theta}{\partial \theta} + u_r\right) + (\Gamma_{03333} + q)\frac{\partial u_z}{\partial z} - \dot{q},$$

$$\dot{T}_{0r\theta} = \Gamma_{01212}\frac{\partial u_\theta}{\partial r} + (\Gamma_{01221} + q)\frac{1}{r}\left(\frac{\partial u_r}{\partial \theta} - u_\theta\right), \quad \dot{T}_{0\theta r} = \Gamma_{02121}\frac{1}{r}\left(\frac{\partial u_r}{\partial \theta} - u_\theta\right) + (\Gamma_{01221} + q)\frac{\partial u_\theta}{\partial r}, \quad (34)$$

$$\dot{T}_{0rz} = (\Gamma_{01331} + q)\frac{\partial u_r}{\partial z} + \Gamma_{01313}\frac{\partial u_z}{\partial r}, \quad \dot{T}_{0zr} = \Gamma_{03131}\frac{\partial u_r}{\partial z} + (\Gamma_{01331} + q)\frac{\partial u_z}{\partial r},$$

$$\dot{T}_{0\theta z} = \Gamma_{02323}\frac{1}{r}\frac{\partial u_z}{\partial \theta} + (\Gamma_{02332} + q)\frac{\partial u_\theta}{\partial z}, \quad \dot{T}_{0z\theta} = \Gamma_{03232}\frac{\partial u_\theta}{\partial z} + (\Gamma_{02332} + q)\frac{1}{r}\frac{\partial u_z}{\partial \theta}$$

where, in general, the instantaneous elastic moduli $\Gamma_{0ijkl}$ and the Lagrange multiplier $q$ depend on the applied mechanical pre-stretches $\lambda_z$ and $\lambda$, where the latter is in turn determined by the applied pressures $p_{in}$ and $p_{ou}$ once the material gradient parameters $\beta_1$ and $\beta_2$ are specified. Consequently, the instantaneous mechanical properties of the FG-EHC can be changed by tailoring the material gradient and adjusting the axial pre-stretch and the applied pressures, which will generate paramount effects on the dynamic behavior of the incremental motions. Specifically, for the strain energy function described by the incompressible Mooney-Rivlin model (21), the nonzero components of the instantaneous elasticity tensors in terms of the principal stretches can be evaluated from Eq. (8) as

$$\Gamma_{01111} = \mu_1\lambda^{-2}\lambda_z^{-2} - 3\mu_2\lambda^2\lambda_z^2, \quad \Gamma_{02222} = \mu_1\lambda^2 - 3\mu_2\lambda^{-2}, \quad \Gamma_{03333} = \mu_1\lambda_z^2 - 3\mu_2\lambda_z^{-2},$$

$$\Gamma_{01212} = \mu_1\lambda^{-2}\lambda_z^{-2} - \mu_2\lambda^{-2}, \quad \Gamma_{02323} = \mu_1\lambda^2 - \mu_2\lambda_z^{-2}, \quad \Gamma_{01313} = \mu_1\lambda^{-2}\lambda_z^{-2} - \mu_2\lambda_z^{-2},$$

$$\Gamma_{02121} = \mu_1\lambda^2 - \mu_2\lambda^2\lambda_z^2, \quad \Gamma_{03232} = \mu_1\lambda_z^2 - \mu_2\lambda^{-2}, \quad \Gamma_{03131} = \mu_1\lambda_z^2 - \mu_2\lambda^2\lambda_z^2, \quad (35)$$

$$\Gamma_{02112} = \Gamma_{1221} = -\mu_2(\lambda^2\lambda_z^2 + \lambda^{-2}), \quad \Gamma_{02332} = \Gamma_{03223} = -\mu_2(\lambda^{-2} + \lambda_z^{-2}),$$

$$\Gamma_{01331} = \Gamma_{03113} = -\mu_2(\lambda^2\lambda_z^2 + \lambda_z^{-2})$$

If we assume that the applied pressures $p_{in}$ and $p_{ou}$ remain unchanged during the incremental motion, the incremental boundary conditions (11) on the inner and outer cylindrical surfaces in the deformed configuration become

$$\dot{T}_{0rr} = p_{in} \frac{\partial u_r}{\partial r}, \ \dot{T}_{0r\theta} = p_{in} \frac{1}{r}\left(\frac{\partial u_r}{\partial \theta} - u_\theta\right), \ \dot{T}_{0rz} = p_{in} \frac{\partial u_r}{\partial z}, \quad (r = a)$$

$$\dot{T}_{0rr} = p_{ou} \frac{\partial u_r}{\partial r}, \ \dot{T}_{0r\theta} = p_{ou} \frac{1}{r}\left(\frac{\partial u_r}{\partial \theta} - u_\theta\right), \ \dot{T}_{0rz} = p_{ou} \frac{\partial u_r}{\partial z}, \quad (r = b)$$

(36)

As described in Section 3, the essential physical variables for the axisymmetric deformation in the pressurized FG-EHC are radially inhomogeneous, leading to the $r$-dependence of the instantaneous elastic moduli. Therefore, the resulting incremental displacement equations of motion, in general, are a system of coupled partial differential equations with variable coefficients. We note that the state-space method (SSM), which combines the state-space formalism with the approximate laminate or multi-layer technique, has been put forward by Chen et al. [32] and Wu et al. [4] to solve similar problems effectively. Specifically, Wu et al. [4] made use of the SSM to investigate the effects of inhomogeneous biasing fields on the circumferential waves in a homogeneous electroelastic tube, Chen et al. [32] employed it to study various bifurcation behaviors of a pressurized FG-EHC subjected to end thrust. The readers who are interested in the SSM are referred to Ding and Chen [55] and Chen and Ding [56] and the references cited therein for more details.

If we choose $\mathbf{Y} = \begin{bmatrix} u_r & u_\theta & u_z & \dot{T}_{0rr} & \dot{T}_{0r\theta} & \dot{T}_{0rz} \end{bmatrix}^{\mathrm{T}}$ as the incremental state vector, the state equation can be readily derived from Eqs. (32)-(34) following a standard way [4,55]. For simplicity, we directly give the final state equation as follows:

$$\frac{\partial \mathbf{Y}}{\partial r} = \mathbf{M}\mathbf{Y}$$

(37)

where $\mathbf{M}$ is the $6 \times 6$ system matrix, with its four partitioned $3 \times 3$ sub-matrices given in Appendix A. Note that the state equation (37) is valid for any form of the strain energy function. When supplemented with the appropriate incremental boundary conditions (36), the state equation (37) can be efficiently solved as shown in the next section. For our purpose in this paper, we have disregarded the output equations which are usually needed for the determination of the physical variables other than the state variables in $\mathbf{Y}$.

## 5. Dispersion relations of axisymmetric waves

Based on the SSM, Wu et al. [4] presented an analysis of the guided circumferential elastic waves (including circumferential SH and Lamb-type waves) in a soft electroactive tube under inhomogeneous electromechanical biasing fields, and investigated in detail the effects of the axial pre-stretch and the radial electric voltage on the circumferential waves. Therefore, the interested reader can be referred to Wu et al. [4] for more details. In addition to the circumferential waves, much effort has been also devoted to the investigation of guided elastic waves propagating along the longitudinal (or axial) direction in (soft) hollow cylinders [14-25,27-30], which is also very important to ultrasonic non-destructive evaluation, signal processing in electronic devices, phononic crystals and metamaterials, etc. [1,3]. For these reasons, in this section, we only consider the small-amplitude axisymmetric wave motions along the longitudinal direction in the pressurized FG-EHC with the underlying axisymmetric deformation determined in Section 3. In this case, the state equation (37) can be simplified considerably as

$$\frac{\partial \mathbf{Y}_k}{\partial r} = \mathbf{M}_k \mathbf{Y}_k, \quad k \in \{1, 2\} \tag{38}$$

where $\mathbf{Y}_1 = \left[ u_\theta, \dot{T}_{0r\theta} \right]^{\mathrm{T}}$, $\mathbf{Y}_2 = \left[ u_r, u_z, \dot{T}_{0rr}, \dot{T}_{0rz} \right]^{\mathrm{T}}$ and

$$\mathbf{M}_1 = \begin{bmatrix} \dfrac{f_1}{r} & \dfrac{1}{\varGamma_{01212}} \\ \rho \dfrac{\partial^2}{\partial t^2} + \dfrac{f_6}{r^2} - \varGamma_{03232} \dfrac{\partial^2}{\partial z^2} & -\dfrac{f_1+1}{r} \end{bmatrix}, \mathbf{M}_2 = \begin{bmatrix} -\dfrac{1}{r} & -\dfrac{\partial}{\partial z} & 0 & 0 \\ -f_2 \dfrac{\partial}{\partial z} & 0 & 0 & \dfrac{1}{\varGamma_{01313}} \\ \rho \dfrac{\partial^2}{\partial t^2} + \dfrac{f_3}{r^2} - f_7 \dfrac{\partial^2}{\partial z^2} & \dfrac{f_4}{r} \dfrac{\partial}{\partial z} & 0 & -f_2 \dfrac{\partial}{\partial z} \\ -\dfrac{f_4}{r} \dfrac{\partial}{\partial z} & \rho \dfrac{\partial^2}{\partial t^2} - f_5 \dfrac{\partial^2}{\partial z^2} & -\dfrac{\partial}{\partial z} & -\dfrac{1}{r} \end{bmatrix} \tag{39}$$

It can be seen from Eqs. (38) and (39) that the six state variables can be divided into two groups of independent physical variables. That is to say, there exist two types of the incremental axisymmetric elastic waves superimposed on the axisymmetric pre-deformation described in Section 3: the T waves described by $\mathbf{Y}_1$ and $\mathbf{M}_1$, with the only mechanical displacement component $u_\theta$; and the L waves associated with $\mathbf{Y}_2$ and $\mathbf{M}_2$, whose nonzero displacement components are $u_r$ and $u_z$.

For time-harmonic axisymmetric waves, we assume

$$\left[ u_r, u_\theta, u_z, \dot{T}_{0rr}, \dot{T}_{0r\theta}, \dot{T}_{0rz} \right] = \left[ aU_r(\xi), aU_\theta(\xi), aU_z(\xi), p_0\varSigma_{0rr}(\xi), p_0\varSigma_{0r\theta}(\xi), p_0\varSigma_{0rz}(\xi) \right] \mathrm{e}^{\mathrm{i}(kz-\omega t)} \tag{40}$$

where $\mathrm{i} = \sqrt{-1}$ is the imaginary unit, $\xi = r/a$ is the dimensionless radial coordinate, $p_0$ denotes a pressure-like quantity having the unit of the elastic modulus N/m², and $k$ and $\omega$ are the axial wave number and circular frequency, respectively.

Substituting Eq. (40) into Eqs. (38) and (39), we have

$$\frac{\mathrm{d}}{\mathrm{d}\xi}\mathbf{V}_k(\xi) = \bar{\mathbf{M}}_k(\xi)\mathbf{V}_k(\xi), \quad k \in \{1,2\} \tag{41}$$

where $\mathbf{V}_1 = \left[U_\theta, \Sigma_{0r\theta}\right]^{\mathrm{T}}$, $\mathbf{V}_2 = \left[U_r, \mathrm{i}U_z, \Sigma_{0rr}, \mathrm{i}\Sigma_{0rz}\right]^{\mathrm{T}}$, and the non-dimensional system matrices $\bar{\mathbf{M}}_1$ and $\bar{\mathbf{M}}_2$ are given by

$$\bar{\mathbf{M}}_1 = \begin{bmatrix} \dfrac{f_1}{\xi} & \dfrac{p_0}{\Gamma_{01212}} \\ -\bar{\Omega}^2 + \dfrac{1}{\xi^2}\dfrac{f_6}{p_0} + \dfrac{\Gamma_{03232}}{p_0}\bar{K}^2 & -\dfrac{f_1+1}{\xi} \end{bmatrix}, \bar{\mathbf{M}}_2 = \begin{bmatrix} -\dfrac{1}{\xi} & -\bar{K} & 0 & 0 \\ f_2\bar{K} & 0 & 0 & \dfrac{p_0}{\Gamma_{01313}} \\ -\bar{\Omega}^2 + \dfrac{1}{\xi^2}\dfrac{f_3}{p_0} + \dfrac{f_7}{p_0}\bar{K}^2 & \dfrac{\bar{K}}{\xi}\dfrac{f_4}{p_0} & 0 & -f_2\bar{K} \\ \dfrac{\bar{K}}{\xi}\dfrac{f_4}{p_0} & -\bar{\Omega}^2 + \dfrac{f_5}{p_0}\bar{K}^2 & \bar{K} & -\dfrac{1}{\xi} \end{bmatrix} \tag{42}$$

in which we have introduced the following dimensionless quantities

$$\bar{\Omega} = \Omega\lambda_a / (\eta-1), \quad \bar{K} = ka = K\lambda_a / (\eta-1) \tag{43}$$

where $\Omega = \omega H \sqrt{\rho / p_0}$ and $K = kH$ are the dimensionless circular frequency and axial wave number, respectively. The dimensionless phase velocity can be defined as $v_p = \Omega / K = c / \sqrt{p_0 / \rho}$, where $c = \omega / k$ is the phase velocity.

It is evident from Eq. (42) that the system matrices $\bar{\mathbf{M}}_k(\xi)$ depend on $\xi$, which makes it difficult to get the exact solution to Eq. (41) directly. For this reason, we will employ the approximate laminate or multi-layer model [57-59] to obtain the approximate analytical solutions. For this purpose, the deformed FG-EHC is equally divided into $n$ thin layers, each with a sufficiently small thickness $h / n$, such that the system matrices $\bar{\mathbf{M}}_k$ within each layer may be assumed approximately as constant rather than variable. Consequently, the solutions in the $j$th layer can be obtained as [55-59]

$$\left.\begin{array}{l} \mathbf{V}_1(\xi) = \exp[(\xi-\xi_{j0})\bar{\mathbf{M}}_{1j}(\xi_{jm})]\mathbf{V}_1(\xi_{j0}) \\ \mathbf{V}_2(\xi) = \exp[(\xi-\xi_{j0})\bar{\mathbf{M}}_{2j}(\xi_{jm})]\mathbf{V}_2(\xi_{j0}) \end{array}\right\}, \quad \left(\xi_{j0} \le \xi \le \xi_{j1}, \quad j=1,2,3,\cdots n\right) \tag{44}$$

where $\bar{\mathbf{M}}_{1j}(\xi_{jm})$ and $\bar{\mathbf{M}}_{2j}(\xi_{jm})$ denote the approximated system matrices for T and L waves, respectively, which are constant within the $j$th layer by taking $\xi = \xi_{jm}$; $\xi_{j0}$, $\xi_{j1}$ and $\xi_{jm}$ are the dimensionless radial coordinates at the inner, outer and middle surfaces of the $j$th layer, respectively, i.e.,

$$\xi_{j0} = 1 + (j-1)\frac{\bar{\eta}-1}{n}, \quad \xi_{j1} = 1 + j\frac{\bar{\eta}-1}{n}, \quad \xi_{jm} = 1 + \frac{(2j-1)(\bar{\eta}-1)}{2n} \tag{45}$$

where we recall that $\bar{\eta} = b/a$. Setting $\xi = \xi_{j1}$ in Eq. (44), we have the relations between the state vectors at the inner and outer surfaces of the $j$th layer as follows

$$\left.\begin{array}{l} \mathbf{V}_1(\xi_{j1}) = \exp[(\bar{\eta}-1)\bar{\mathbf{M}}_{1j}/n]\mathbf{V}_1(\xi_{j0}) \\ \mathbf{V}_2(\xi_{j1}) = \exp[(\bar{\eta}-1)\bar{\mathbf{M}}_{2j}/n]\mathbf{V}_2(\xi_{j0}) \end{array}\right\} \tag{46}$$

Making use of the continuity conditions at each fictitious interface between the thin layers, we obtain from Eq. (46)

$$\mathbf{V}_k(\bar{\eta}) = \mathbf{S}_k\mathbf{V}_k(1), \quad k \in \{1, 2\} \tag{47}$$

where $\mathbf{S}_k = \prod_{j=n}^{1}\exp[(\bar{\eta}-1)\bar{\mathbf{M}}_{kj}/n]$ are the global transfer matrices of second-order ($k=1$) or fourth-order ($k=2$), through which the state vectors at the outer surface $\mathbf{V}_k(\bar{\eta})$ are directly related to those at the inner surface $\mathbf{V}_k(1)$.

For the incremental axisymmetric wave propagation in the pressurized FG-EHC, the incremental boundary conditions (36) on the inner and outer cylindrical surfaces in the deformed configuration reduce to

$$\dot{T}_{0rr} = -p_{in}\left(\frac{u_r}{r} + \frac{\partial u_z}{\partial z}\right), \ \dot{T}_{0r\theta} = -p_{in}\frac{u_\theta}{r}, \ \dot{T}_{0rz} = p_{in}\frac{\partial u_r}{\partial z}, \quad (r = a)$$

$$\dot{T}_{0rr} = -p_{ou}\left(\frac{u_r}{r} + \frac{\partial u_z}{\partial z}\right), \ \dot{T}_{0r\theta} = -p_{ou}\frac{u_\theta}{r}, \ \dot{T}_{0rz} = p_{ou}\frac{\partial u_r}{\partial z}, \quad (r = b) \tag{48}$$

where we have utilized $\partial/\partial\theta = 0$ as well as the incremental incompressibility constraint (32).

Substituting Eq. (40) into Eq. (48) and using Eq. (43)$_2$, we have the dimensionless form of the incremental boundary conditions (48) as follows:

$$\Sigma_{0rr}(1) = -\bar{p}_{in}\left[U_r(1) + iU_z(1)\bar{K}\right], \ \Sigma_{0r\theta}(1) = -\bar{p}_{in}U_\theta(1), \ i\Sigma_{0rz}(1) = -\bar{p}_{in}\bar{K}U_r(1),$$

$$\Sigma_{0rr}(\bar{\eta}) = -\bar{p}_{ou}\left[\bar{\eta}^{-1}U_r(\bar{\eta}) + iU_z(\bar{\eta})\bar{K}\right], \ \Sigma_{0r\theta}(\bar{\eta}) = -\bar{p}_{ou}\bar{\eta}^{-1}U_\theta(\bar{\eta}), \ i\Sigma_{0rz}(\bar{\eta}) = -\bar{p}_{ou}\bar{K}U_r(\bar{\eta}) \tag{49}$$

where $\bar{p}_{in} = p_{in} / p_0$ and $\bar{p}_{ou} = p_{ou} / p_0$ are the dimensionless pressures. By applying the incremental boundary conditions (49) to Eq. (47), we acquire two sets of the linear algebraic equations as

$$\mathbf{D}_1 \left[ U_\theta(1), U_\theta(\bar{\eta}) \right]^{\mathrm{T}} = \mathbf{0}, \quad \mathbf{D}_2 \left[ U_r(1), \mathrm{i} U_z(1), U_r(\bar{\eta}), \mathrm{i} U_z(\bar{\eta}) \right]^{\mathrm{T}} = \mathbf{0} \tag{50}$$

where $\mathbf{D}_1$ and $\mathbf{D}_2$ are the coefficient matrices of second-order and fourth-order, respectively, whose nonzero elements are given as

$$D_{1k1} = S_{1k1} - \bar{p}_{in} S_{1k2}, \quad D_{112} = -1, \quad D_{122} = \bar{p}_{ou} \bar{\eta}^{-1}, \quad (k = 1, 2) \tag{51}$$

and

$$D_{2k1} = S_{2k1} - \bar{p}_{in} \left( S_{2k3} + \bar{K} S_{2k4} \right), \quad D_{2k2} = S_{2k2} - \bar{p}_{in} \bar{K} S_{2k3}, \quad (k = 1, 2, \cdots, 4)$$
$$D_{213} = -1, \quad D_{224} = -1, \quad D_{233} = \bar{p}_{ou} \bar{\eta}^{-1}, \quad D_{234} = D_{243} = \bar{K} \bar{p}_{ou} \tag{52}$$

Note that $S_{kij}$ are the elements of the global transfer matrices $\mathbf{S}_k$. Since Eq. (50) shall have non-trivial solutions, the determinants of the coefficient matrices must vanish, i.e.,

$$\left| \mathbf{D}_k \right| = 0, \quad k \in \{1, 2\} \tag{53}$$

which determine the dispersion relations between the dimensionless axial wave number $K$ and the dimensionless circular frequency $\Omega$ for the T and L waves, respectively.

Furthermore, when there is no external pressure $p_{ou} = 0$, Eqs. (50) and (53) can be simplified to

$$\left( S_{121} - \bar{p}_{in} S_{122} \right) U_\theta(1) = 0, \quad \mathbf{D}_2^{in} \left[ U_r(1), \mathrm{i} U_z(1) \right]^{\mathrm{T}} = \mathbf{0} \tag{54}$$

and

$$S_{121} - \bar{p}_{in} S_{122} = 0, \quad \left| \mathbf{D}_2^{in} \right| = 0 \tag{55}$$

where the nonzero elements of the coefficient matrix $\mathbf{D}_2^{in}$ are given as

$$D_{2j1}^{in} = S_{2k1} - \bar{p}_{in} \left( S_{2k3} + \bar{K} S_{2k4} \right), \quad D_{2j2}^{in} = S_{2k2} - \bar{p}_{in} \bar{K} S_{2k3}, \quad (j = k - 2; \ k = 3, 4) \tag{56}$$

For the case that the internal pressure vanishes, i.e., $p_{in} = 0$, we can derive the simplified dispersion relations similarly.

## 6. Numerical results and discussions

In this section, in order to investigate the combined effects of the material gradient and the mechanical biasing field (induced by the applied pressure and/or the axial pre-stretch) on the nonlinear response and the dispersion relations for the T and L waves, numerical calculations are conducted for an FG-EHC made of an incompressible isotropic Mooney-Rivlin material, whose strain energy function is characterized by Eqs. (21)-(23). In the numerical computations, we always set $\mu_{10} = 1.858 \times 10^5 \, \text{Pa}$ and $\mu_{20} = -0.1935 \times 10^5 \, \text{Pa}$, which were used by Batra et al. [60] for the rubber they considered. For simplicity, the two material gradient parameters $\beta_1$ and $\beta_2$ are set to be equal, i.e., $\beta = \beta_1 = \beta_2$. Furthermore, the shear modulus $\mu(R) = \mu_1(R) - \mu_2(R)$ of the Mooney-Rivlin material for an infinitesimal deformation should be positive, which requires that the permissible material gradient parameter should satisfy $\beta > -1/\eta$. In addition, the initial shape factor $\eta$, i.e., the ratio of the outer radius to the inner radius of the undeformed hollow cylinder, is all fixed as $\eta = B/A = 2$. Then we have $\beta > -0.5$. Finally, the pressure-like quantity $p_0$ is taken to be equal to the material constant $\mu_{10}$ in the following numerical evaluations.

### 6.1. Nonlinear static response

Before studying the effects of the material gradient and the mechanical biasing field on the axisymmetric wave propagation characteristics in a pressurized FG-EHC in Section 6.2, we first show the nonlinear static response of an FG-EHC to different mechanical biasing fields. Utilizing the results presented in this subsection, we can clearly reveal some unique correlations between the frequency of the axisymmetric waves and the material gradient as well as the mechanical biasing field.

In Fig. 2 we plot $\lambda_a$ based on Eq. (26) as a function of the dimensionless pressure difference $\Delta p^* = \Delta p / \mu_{10}$ for an FG-EHC using the incompressible Mooney-Rivlin model (21)-(23) for different values of the material gradient parameter $\beta$ and three different axial pre-stretches $\lambda_z$. In our numerical calculations, the lower bound of $\lambda_a$ is consistently set to be 0.5, which corresponds to a positive value of the pressure difference for the three axial pre-stretches and all material gradient parameters used in Fig. 2. Note that, for a given material gradient, an ever bigger positive pressure difference leading to a smaller $\lambda_a$ may make the FG-EHC mechanically instable [32]. It was described in Section 3 that, when there is no pressure applied to the inner and outer surfaces, a uniform deformation state always exists even if the EHC subjected to an axial tension or compression is inhomogeneous in the radial direction [32]. Thus the circumferential stretch $\lambda_a$ at the inner surface may be expressed by $\lambda_a = \lambda_z^{-1/2}$ for $\Delta p^* = 0$. Specifically, the circumferential stretches are $\lambda_a = 1.11803$, 1 and 0.70711 for the pre-stretches $\lambda_z = 0.8$, 1, and 2, respectively, as shown in Fig. 2. In addition, it is seen from Fig. 2 that, for a fixed pre-stretch and material gradient, there is a critical negative pressure difference $\Delta p_c^* < 0$, beneath which no solution of the axisymmetric deformation exists and the FG-EHC collapses. For example, when $\beta = 5$, the critical pressure differences are $\Delta p_c^* = $ -7.42, -6.20, and -4.04 for $\lambda_z = 0.8$, 1, and 2, respectively, as indicated by the vertical lines in Fig. 2. This phenomenon is in accordance with the findings by Melnikov and Ogden [61] on finite deformations of a pressurized cylindrical tube of the soft dielectric material subjected to electromechanical biasing fields. Consequently, the pressure difference should be appropriately selected in order to ensure the existence of the axisymmetric deformation state in the FG-EHC. Furthermore, the critical negative pressure difference depends on the axial pre-stretch and the material gradient. To be more specific, $\Delta p_c^*$ decreases with the material gradient $\beta$ for a given pre-stretch, while for a specified material gradient, raising the pre-stretch $\lambda_z$ increases the value of $\Delta p_c^*$. This indicates that an increase in the material gradient and axial compression can enlarge the existence scope of the axisymmetric deformation and enhance the adjustment range of the pressure difference for the FG-EHC. It is also noted from Fig. 2 that, the absolute value of $\Delta p^*$ which is required to reach the same level of $\lambda_a$ increases with the material gradient $\beta$ independent of whether the pressure difference $\Delta p^*$ is positive or negative. Therefore, an increase in the material gradient stiffens the FG-EHC.

### 6.2 Wave propagation analysis

### 6.2.1 Validation of the proposed approach

In this subsection, the proposed approach based on the state-space formalism along with the approximate laminate technique (hereafter abbreviated as SSM) will be validated with respect to its accuracy and convergence for investigating the axisymmetric waves in the EHC. It should be emphasized once more that the T waves denote the incremental torsional waves, while the L waves represent the incremental longitudinal waves. These two types of incremental axisymmetric waves are uncoupled, as described in Section 5. The material gradient and the mechanical biasing field considered in the paper will not influence this decoupling feature. It is noted here that the uncoupled T and L waves can be derived from the incremental *non-axisymmetric* waves in Appendix B by taking the circumferential wave number $m = 0$ [63,64]. The investigation of non-axisymmetric waves ($m \neq 0$) will go beyond the scope of the present study, and will be thus not considered in the following.

If there is no pressure applied to a homogeneous EHC, its deformation subjected to a pre-stretch only will be homogeneous, as mentioned earlier. It is noted that Su et al. [62] has analytically derived the dispersion relations of non-axisymmetric waves in an infinite soft electroactive hollow cylinder under a uniform deformation. Thus, exact dispersion relations of the T and L waves considered in this paper can also be deduced for a homogeneous EHC from those in Su et al. [62] through a proper degenerate analysis. Consequently, the accuracy of SSM may be verified by making a comparison of the numerical results with these exact solutions, which are omitted here for brevity. In Fig. 3, we compare the frequency spectra ($\Omega - K$ curves) of the first five wave modes obtained by the exact solutions [62] and SSM for a homogeneous EHC described by a Mooney-Rivlin model at three different pre-stretches $\lambda_z = 0.8$, 1 and 2. The frequency spectra for the T waves are plotted in Fig. 3(a) while those for the L waves are displayed in Fig. 3(b). Note that the lines correspond to the exact branches while the markers to those of SSM, for which the homogeneous hollow cylinder is divided into 30 thin layers. It can be seen from Fig. 3 that the SSM results agree very well with the exact solutions in the entire wave number range for both T and L waves, which validates the accuracy of the SSM.

As shown in Fig. 3, the frequency of the T waves increases considerably with the pre-stretch in the entire wave number range, while that of the L waves increases with the pre-stretch in most of the wave number range except in the low wave number region for the higher modes than the second mode. In addition, it is observed from Fig. 3(a) that, the first branch of the frequency spectra of T waves always passes through the origin and is always a straight line with a slope $\sqrt{\lambda_z \left( \lambda_z - \mu_{20} / \mu_{10} \right)}$, indicating that the fundamental mode is always nondispersive. This phenomenon has been observed by Shearer et al. [28] but left out by Engin and Suhubi [27] and Su et al. [62]. Besides, the first branch of the L waves at large wave numbers is almost a straight line. In fact, the first modes of the T and L waves at a large wave number asymptotically approach the modified shear wave and the modified Rayleigh surface wave, respectively, in a pre-stretched homogeneous EHC.

In addition, we also compute the displacement variations of the first four L wave modes in the long wavelength limit, specifically at $K = 0$, and the results indicate that these elastic waves correspond to the uniform extensional mode (with zero cut-off frequency), the first radial breathing mode, and the first and second axial-shear modes, respectively [64,65]. The axial displacements of the two axial-shear modes have antisymmetric- and symmetric-like variations with respect to the mid-surface of the hollow cylinder. Furthermore, the curves of the first three nonzero dimensionless cut-off frequencies $\Omega_{co}$ at $K = 0$ of the T and L waves versus the axial pre-stretch in a homogeneous EHC without pressures are displayed in Figs. 4(a) and 4(b), respectively. The uniform extensional mode of the L waves is thus excluded there due to its zero cut-off frequency. It can be found from Fig. 4 that all cut-off frequencies of the T waves increase monotonically with the pre-stretch. On the other hand, the second and third cut-off frequencies of the L waves decrease continuously, while the first one increases with the pre-stretch. That is to say, the pre-stretch raises the cut-off frequency of the first radial breathing mode but lowers those of the first and second axial-shear modes.

For a pressurized FG-EHC, it is intractable to obtain exact dispersion relations owing to the radially inhomogeneous biasing field. However, the SSM can be utilized to calculate the frequency spectra, which may be verified through checking its convergence behavior. At a wave number $K = 2.6$, the variations of the second frequency for the T waves and the first frequency for the L waves with the number of the discretized thin layers for a pressurized FG-EHC with $\lambda_z = 1$ are displayed in Fig. 5 for different material gradients and pressure differences. It can be observed from Fig. 5 that when the thin layer number increases, the frequencies for all cases asymptotically approach specific values, which demonstrates that the SSM has an excellent convergence characteristic. As pointed out by Wu et al. [4], the approximate laminate model will gradually approach the original FG-EHC that is radially inhomogeneous due to the pressure difference when the layer number increases. As a consequence, we can use the present SSM to obtain accurate numerical results with an arbitrary precision. In the following calculations, the number of the equally discretized thin layers is taken to be 30 for which the numerical results can be considered to be highly accurate, as illustrated in Fig. 5.

Before using the SSM to study the incremental wave propagation characteristics in a pressurized FG-EHC, we first investigate how the frequency spectra vary with the internal or external pressure for a fixed pressure difference. The first five branches of the frequency spectra for both T and L waves (with $\Delta p^* = 0$ and $\Delta p^* = -1$, respectively) are shown in Fig. 6 for a pressurized FG-EHC with $\beta = 5$ and $\lambda_z = 1$ at different values of the internal pressure. We note that, for a fixed pressure difference, the frequency spectra are not affected by the internal or external pressure although both the internal and external pressures are involved independently in Eq. (52). It means that it is the pressure difference, not the internal or external pressure separately, that determines the frequency spectra. This phenomenon exactly agrees well with the analytical conclusions drawn upon the conventional displacement method presented in Appendix B. Therefore, we only need to specify the pressure difference in the following numerical calculations, while the internal or external pressure can be assigned with an arbitrary value.

*6.2.2 Results for T waves*

After validating the proposed approach by verifying its convergence and accuracy in the previous subsection, we now utilize it to obtain the frequency spectra for the T waves propagating in an FG-EHC subjected to an axial pre-stretch and a radial pressure difference.

*6.2.2.1 Effect of the pressure difference*

For three values of the material gradient parameter $\beta = 0$, 5, and 10, the first five branches of the frequency spectra for the T waves are depicted in Figs. 7(a)-7(c), respectively, for a pressurized FG-EHC with $\lambda_z = 1$ at different values of the pressure difference. Particularly, the results for $\Delta p^* = 0$ in Figs. 7(a)-7(c) correspond to the case of the FG-EHC subjected to purely mechanical pre-stretch. As previously mentioned, in order to ensure the existence of the solution for the axisymmetric deformation and avoid the instability or bifurcation of the hollow cylinder under the pressure difference, there exists an allowable range of the pressure difference, which depends on the axial pre-stretch and the material gradient. Specifically, it can be seen from Fig. 2 that, the allowable pressure differences are (-0.752, 2.180), (-6.166, 15.307), and (-11.579, 28.434) for $\lambda_z = 1$ and $\beta = 0$, 5, and 10, respectively. Therefore, the values of the pressure difference in Fig. 7 are all in the allowable range.

For the homogeneous EHC with $\beta = 0$, it can be found from Fig. 7(a) that the frequency increases slightly with the pressure difference in the entire wave number range. We can also observe that the fundamental mode for a non-zero pressure difference also passes through the origin and is non-dispersive in the entire range. The increase of the pressure difference raises the curve slope of the fundamental mode, i.e., the group velocity. These phenomena are analogous to those observed by Shearer et al. [28]. In addition, the pressure difference hardly changes the curve slopes of the higher-order modes. For the FG-EHC shown in Figs. 7(b) and 7(c), qualitatively similar phenomena to those in Fig. 7(a) can be found except that the fundamental mode becomes dispersive, even for $\Delta p^* = 0$, which is physically reasonable because of the material inhomogeneity of the FG-EHC.

As shown in Fig. 7, the frequency of the T waves increases with the pressure difference for a fixed material gradient. Therefore, in order to clearly reveal how the frequency varies with the pressure difference, the variations of the frequency of the T wave modes with the pressure difference for an FG-EHC with $\lambda_z = 1$ are illustrated in Fig. 8 for two specific material gradient parameters $\beta = 0$ (Figs. 8(a) and 8(c)) and $\beta = 5$ (Figs. 8(b) and 8(d)). Figs. 8(a) and 8(b) show the first three nonzero cut-off frequencies at $K = 0$, while the first four frequencies at $K = 5$ are displayed in Figs. 8(c) and 8(d). It should be emphasized again that the allowable ranges of the pressure difference for different material gradients are different. It can be noted from Fig. 8 that the frequencies for both wave numbers and all modes indeed increase monotonically with the applied pressure difference for a specified material gradient. Besides, the tunable frequency range for the same wave number and mode is enlarged when the material gradient increases. The results for other pre-stretches are essentially the same as those for $\lambda_z = 1$ and hence are omitted here. To conclude, an increase in the material gradient expands the tunable range of the frequency of the T waves and enhances the ability for adjusting elastic waves in the FG-EHC by the pressure difference.

At small and large wave numbers ($K = 0.2$ and $K = 10$, respectively), the normalized displacement variations $U_\theta^*$ of the lowest torsional mode along the dimensionless radial coordinate $\xi$ with $\lambda_z = 1$ are calculated and plotted in Figs. 9(a) and 9(b) for different values of $\Delta p^*$ at $\beta = 0$ and $\beta = 5$, respectively. It is worth mentioning that the displacement amplitude is normalized to be within $\pm 1$ by its absolute maximum value for each combination of the pressure difference and material gradient. The end-point of the dimensionless radial coordinate becomes different when the pressure difference changes. As shown in Fig. 9, in the long wave case ($K = 0.2$), the displacements show a linear variation along the radial coordinate for all considered pressure differences and material gradients, which is reasonable recalling the same behavior of a solid rod [7]. However, in the limit of short waves ($K = 10$), the mode shapes are nonlinear except for $\Delta p^* = \beta = 0$. More specifically, the displacement amplitude at the outer surface for $\beta = 0$ and $\Delta p^* > 0$ is larger than that at the inner surface, while the situation for $\beta = 0$ and $\Delta p^* < 0$ exhibits an opposite trend. For $\beta = 5$, the mode shape is nearly independent of the pressure difference and the torsional mode is mainly localized on the inner surface. In a word, the mode shape in the limit of short waves may be well utilized to characterize the material gradient behavior and the mechanical biasing field.

*6.2.2.2 Effect of the material gradient*

Now our interest is turned to determine how the material gradient influences the propagation of the T waves in a pressurized FG-EHC. In Fig. 10, the frequency spectra for the T waves are depicted for a pressurized FG-EHC with $\lambda_z = 1$ at different values of the material gradient parameter for three specified pressure differences $\Delta p^* = -2$, 0, and 2. Compared with the effect of the pressure difference displayed in Fig. 7, the material gradient significantly increases the frequencies and the curve slopes of all modes (i.e., the group velocities) in the entire wave number range.

Furthermore, the variations of the frequency of the T wave modes with the material gradient parameter are shown in Fig. 11 for different values of the pressure difference and two given pre-stretches $\lambda_z = 1$ (Figs. 11(a) and 11(c)) and $\lambda_z = 2$ (Figs. 11(b) and 11(d)). The first four nonzero cut-off frequencies at $K = 0$ and the first five frequencies at $K = 5$ are plotted in Figs. 11(a), (b) and Figs. 11(c), (d), respectively. It can be observed from Fig. 2 that the material gradient can take any large value up to infinity, but its allowable lower bound depends on the pre-stretch and the pressure difference. As we know, the positiveness of the shear modulus of the undeformed state requires $\beta > -0.5$, and we denote the allowable range of the pressure difference for $\beta = -0.5$ as $(\Delta p_1^*, \Delta p_2^*)$. This range can be readily obtained by taking $\beta \to -0.5$ in Fig. 2. It is noticed however that the actual lower bound of the material gradient may be larger than $-0.5$, which is determined by the existence of the axisymmetric deformation for $\Delta p^* < \Delta p_1^*$ or by the constraint of $\lambda_a = 0.5$ (a prerequisite in our numerical calculations as mentioned earlier) for $\Delta p^* > \Delta p_2^*$. It can be seen from Fig. 11 that, when the pressure difference and the pre-stretch are given, the frequencies for both wave numbers and all modes indeed increase significantly and monotonically with the material gradient, and additionally, the gaps between two neighboring modes get larger with the material gradient increasing in the allowable range. Besides, the frequency has a slight rise when the pressure difference goes up, which is in accordance with the result observed in Fig. 7.

At small and large wave numbers ($K = 0.2$ and $K = 10$, respectively), we also plot the normalized displacement variations $U_\theta^*$ of the lowest torsional mode with $\lambda_z = 1$ in Fig. 12 for different values of $\beta$ at $\Delta p^* = 0$ and $\Delta p^* = -2$. It is noted that the end-point of the dimensionless radial coordinate for $\Delta p^* = 0$ keeps unchanged since the ratio of the outer radius to the inner one does not change before and after the deformation as described in Section 3. As shown in Fig. 12, for fixed pressure difference and wave number, the mode shape is essentially independent of the material gradient. However, for a fixed material gradient, the mode variations are different for different wave numbers except for $\beta = 0$.

In a word, the nonlinear and monotonic dependence of the frequency on the material gradient and the difference in mode shapes at different pressure differences provide a possibility to characterize the material properties and the pre-stress state or detect the structural defects such as cracks of the pressurized FG-EHC via in-situ ultrasonic nondestructive evaluation. In addition, for a given working condition (i.e., the pre-stretch and the pressure difference are known), we can readily adjust or optimize the wave propagation characteristics in an FG-EHC via tailoring its material compositions.

### 6.2.3 Results for L waves

Having discussed the effects of the material gradient, pressure difference and axial pre-stretch on the T wave propagation characteristics in the previous subsection, now our attention focuses on how these factors affect the propagation of the L waves in the pressurized FG-EHC. The values of the material gradient, pressure difference and axial pre-stretch under consideration are assumed to be the same as those adopted in the previous subsection for the T waves.

*6.2.3.1 Effect of the material gradient*

For the pre-stretch $\lambda_z = 1$ and three different pressure differences $\Delta p^* = -2$, 0, and 2, the frequency spectra and the phase velocity spectra ($v_p - K$ curves) for the L waves are presented in Fig. 13 for a pressurized FG-EHC at different values of the material gradient parameter. In particular, the results for $\Delta p^* = 0$ and $\beta = 0$ corresponding to the homogeneous EHC without pressure difference and pre-stretch are shown in Figs. 13(b) and 13(e). It is seen that the fundamental mode is dispersive unlike the T waves considered in the previous subsection and there is no cutoff frequency for this mode. In this case, the phase velocity of the fundamental mode in the limit of long waves ($K \to 0$) approaches $v_p = 1.82$, i.e., $c = 1.82\sqrt{\mu_{10}/\rho} = \sqrt{3}c_T$, where $c_T = \sqrt{(\mu_{10} - \mu_{20})/\rho}$ is the transverse wave velocity in the undeformed state. This result coincides with the classical result for a solid rod [29,62,63]. In the limit of short waves ($K \to \infty$), the phase velocity of the fundamental mode tends to $v_p = 1.004$ (i.e., $c = 1.004\sqrt{\mu_{10}/\rho} = 0.955c_T$) which is the Rayleigh surface wave velocity in an elastic half-space [7,29].

Additionally, we observe from Fig. 13 obvious effects of the material gradient on the frequency spectra as well as the phase velocity spectra. Specifically, for the considered cases, the frequency and phase velocity increase dramatically with the material gradient in the entire wave number range, especially at large wave numbers, and the material gradient also raises noticeably the curve slope or the group velocity of all modes. These phenomena for the L waves are pretty similar to those for the T waves. Furthermore, all the modes are dispersive in all cases and the phase velocities of the higher-order modes start from infinity with a finite cutoff frequency depending on the material gradient and the pressure difference. However, the phase velocities of the fundamental mode all start from a finite value which is also related to the material gradient and the pressure difference. Particularly, the phase velocities of the fundamental mode in all cases exhibit a non-monotonic behavior that the phase velocities first have a rapid decrease at small wave numbers, then arrive at a minimum, and subsequently increase gradually with the wave number, asymptotically approaching the modified Rayleigh surface wave velocities. This trend will become more evident for a larger material gradient. A similar observation has been obtained and explained by Shmuel and deBotton [29] for the axisymmetric wave propagation in soft electroactive tubes subjected to a radially applied electric field and an axial mechanical load. This non-monotonic variation of the fundamental mode should be attributed to the complex wave interaction with the geometric boundaries in terms of the thickness and mean radius of the deformed FG-EHC [29,62].

Fig. 14 displays the dimensionless frequencies of the L wave modes as functions of the material gradient parameter for different pressure differences and two different pre-stretches $\lambda_z = 1$ (Figs. 14(a) and 14(c)) and $\lambda_z = 2$ (Figs. 14(b) and 14(d)). The first four nonzero cut-off frequencies at $K = 0$ and the first five frequencies at $K = 5$ are shown in Figs. 14(a), (b) and Figs. 14(c), (d), respectively. As mentioned in the previous subsection on the T waves, the allowable lower bound of the material gradient depends on the pre-stretch and the pressure difference. It can be found from Fig. 14 that, for a fixed pre-stretch and a specified pressure difference $\Delta p^* \geq 0$, the frequencies for both wave numbers and all modes indeed increase dramatically and monotonically with the material gradient, whereas for a pressure difference $\Delta p^* < 0$, the frequencies exhibit a non-monotonic variation with the material gradient except the first nonzero cut-off frequency corresponding to the first radial breathing mode at $K = 0$, which also increases continuously with the material gradient. The other frequencies for $\Delta p^* < 0$ first become smaller when the material gradient decreases, and then increase with the material gradient declining towards the allowable lower bound. The reason for these phenomena can be explained as follows. When the material gradient is far away from the lower bound regardless of the sign of the pressure difference, a decrease in the material gradient alters slightly the circumferential stretch as shown in Fig. 2. Therefore, the frequency variation trend is up to the variation of the material gradient, i.e., the frequency reduces at this moment with the decrease of the material gradient leading to the decrease of the stiffness of the FG-EHC. However, as the material gradient approaches the allowable lower bound for $\Delta p^* < 0$, a slight declining of the material gradient will result in a remarkable increase of the circumferential stretch, as also can be seen from Fig. 2, which has a tendency to making the FG-EHC stiffer. Consequently, the increase of the frequency due to a considerable increase in the nonlinear deformation surpasses the decrease amount of the frequency owing to the small drop in the material gradient. As a result, the frequency increases reversely with the decrease of the material gradient. On the other hand, as the material gradient tends to the lower bound for $\Delta p^* > 0$. the variation of the material gradient changes the circumferential stretch a little as depicted in Fig. 2, and hence the frequency will still decrease when the material gradient goes down. In addition, the gaps between two neighboring modes become larger when the material gradient increases in the allowable range, which is the same as that for the T waves. Therefore, analogous to the T waves, the nonlinear changes in the frequency of the L waves with the material gradient can be used to conduct on-line ultrasonic nondestructive testing for the property and pre-stress characterization or the structural defect and crack detection of the pressurized FG-EHC. As mentioned previously, by varying the material composition that leads to a change in the material gradient, the wave propagation behavior in an FG-EHC for a known working condition can be tuned and optimized.

At small and large wave numbers, i.e., $K = 0.2$ (Figs. 15(a) and 15(c)) and $K = 10$ (Figs. 15(b) and 15(d)) respectively, the variations of the normalized radial and axial displacement amplitudes $U_r^*$ and $U_z^*$ of the lowest L wave mode are plotted for $\lambda_z = 1$ and different values of $\beta$ at $\Delta p^* = 0$ (Figs. 15(a) and 15(b)) and $\Delta p^* = -2$ (Figs. 15(c) and 15(d)). Note here that the mode shapes are normalized by the maximum absolute value among the two displacement components. In our numerical calculations, the normalized amplitudes of the radial and axial displacements are real and pure imaginary, respectively, indicating that they are always 90-degrees out of phase, which can also be seen from the incremental state vector. It can be clearly found from Fig. 15 that, at the small wave number $K = 0.2$, the radial displacement almost vanishes, while the axial one predominates and has a uniform variation for all material gradients and pressure differences. This observation indicates that the lowest L wave mode in the limit of long waves corresponds to the uniform extensional mode as described in Subsection 6.2.1. In the limit of short waves $K = 10$, the lowest L wave mode behaves like Rayleigh-type surface waves either on the outer or inner surface of the deformed FG-EHC, and the radial displacement prevails. Specifically, the maximum displacement amplitude for $\Delta p^* = 0$ appears at the outer and inner surfaces for the material gradients $\beta = 2$ and $\beta = 5$, respectively. For $\Delta p^* = -2$, the maximum displacement amplitude is achieved on the inner surface for both $\beta = 2$ and $\beta = 5$.

*6.2.3.2 Effect of the pressure difference*

For the pre-stretch $\lambda_z = 1$ and three different material gradient parameters $\beta = 0$, 5, and 10, the frequency spectra and the phase velocity spectra for the first three branches of the L waves are illustrated in Fig. 16 for a pressurized FG-EHC at different values of the pressure difference. It can be observed from Fig. 16 that the frequency spectra for all modes and the phase velocity spectra for higher-order modes have a monotonous behavior in all cases. However, the phase velocities of the fundamental mode for all cases first decrease rapidly at small wave numbers, then increase gradually with the wave number, and finally asymptotically approach the modified Rayleigh surface wave velocities. These phenomena are independent of the pressure difference and the material gradient, and are qualitatively similar to those observed in Fig. 13. On the other hand, differing substantially from the T waves considered in Section 6.2.2, the effect of the pressure difference on the L wave propagation characteristics is more complex, which will be discussed in detail below.

To highlight the role of the pressure difference in the L wave propagation, the variations of the first two nonzero dimensionless cut-off frequencies of the L wave modes at $K = 0$ with the pressure difference $\Delta p^*$ are shown in Fig. 17 for two pre-stretches $\lambda_z = 1$, 2 and two material gradient parameters $\beta = 0$, 5. It can be seen that the first cut-off frequency corresponding to the first radial breathing mode keeps increasing with the pressure difference, while the second one associated with the first axial-shear mode has a continuous drop. Another interesting phenomenon in Fig. 17 is that, after a certain pressure difference, for example $\Delta p^* \approx 1.5$ at $\beta = 0$, the frequency of the radial mode is larger than that of the axial-shear mode, which means physically that the interaction between the two modes becomes substantial and the wave mode shapes change rapidly around the corresponding pressure difference.

In addition, Figs. 18 and 19 illustrate the dimensionless frequencies of the first three L wave modes at two representative wave number values $K = 2$ and $K = 8$, respectively, as functions of the pressure difference for two different pre-stretches $\lambda_z = 1$ and $\lambda_z = 2$ at three different material gradient parameters $\beta = 0$, 2, and 5. As described above, in order to ensure the axisymmetric deformation and avoid the instability or bifurcation, there exist different allowable ranges of the pressure difference for different axial pre-stretches and material gradients. Specifically, as can be seen from Fig. 2, for a fixed material gradient, the allowable range of the pressure difference decreases with the pre-stretch, whereas for a fixed pre-stretch, the allowable range is enlarged as the material gradient increases. It can be observed from Figs. 18 and 19 that, for both wave numbers, the frequencies for all modes have a considerable rise when the pre-stretch increases for a fixed pressure difference. In addition, similar to the T waves shown in Fig. 8, the tunable frequency range of the same wave number and mode for the L waves becomes larger when the material gradient increases. Therefore, these results suggest the use of the FG-EHC as a tunable waveguide via material tailoring in combination with the adjustment of the pre-stretch and the pressure difference.

For $K = 2$ shown in Fig. 18 and a specified material gradient, the frequency of the fundamental mode decreases monotonically with the allowable pressure difference for both pre-stretches $\lambda_z = 1$ and $\lambda_z = 2$. However, the frequencies of the higher-order modes exhibit a non-monotonic behavior for both pre-stretches. In this case, the frequency first goes down to a minimum near the lower bound of the allowable range corresponding to a sharp inflation as shown in Fig. 2, and then increases gradually when the pressure difference increases in the allowable range. Fig. 19 displays the results corresponding to $K = 8$. For the pre-stretch $\lambda_z = 1$, the frequencies of the first three modes all decrease monotonically with the allowable pressure difference for three given material gradient parameters. Nonetheless, the situation for the pre-stretch $\lambda_z = 2$ is different from that for $\lambda_z = 1$. Specifically, the frequency of the fundamental mode still declines continually with the pressure difference, while the frequencies of the higher-order modes first go down rapidly near the lower bound of the allowable range, and then almost remain unchanged with the pressure difference increasing.

In brief, the intricate dependence on the pressure difference of the L wave propagation characteristics in the FG-EHC is a consequence of multiple factors, including the wave number range, the wave mode, the material gradient, and the axial pre-stretch.

Finally, for small and large wave numbers, i.e., $K = 0.2$ (Figs. 20(a) and 20(c)) and $K = 10$ (Figs. 20(b) and 20(d)) respectively, we also display the variations of the normalized radial and axial displacement amplitudes $U_r^*$ and $U_z^*$ of the lowest L wave mode for $\lambda_z = 1$ and different values of $\Delta p^*$ at $\beta = 0$ (Figs. 20(a) and 20(b)) and $\beta = 5$ (Figs. 20(c) and 20(d)). The results are quite similar to that in the Subsection 6.2.3.1, and their discussion is thus omitted here for brevity. The only issue which should be emphasized is that the pressure difference significantly alters the mode shapes in the limit of short waves at $K = 10$ (see Figs. 20(b) and 20(d)).

## 7. Conclusions

In this paper, we examined the axisymmetric guided waves, including the T and L waves, propagating in a pressurized FG-EHC. Firstly, based on the nonlinear elasticity theory, we addressed the axisymmetric deformation of the FG-EHC subjected to an axial pre-stretch and a radial pressure difference for a general strain energy function, and obtained the explicit expressions of the nonlinear response and the radially inhomogeneous physical variables in the pressurized FG-EHC for the Mooney-Rivlin material model with radial affine variations of the involved parameters. Secondly, using the linear incremental theory, we derived the state-space formalism for the incremental fields in cylindrical coordinates without a specification of the strain energy function. Thirdly, employing a solution approach which combines the state-space formalism and the approximate laminate technique, we obtained the dispersion relations for both T and L waves in the pressurized FG-EHC with radially inhomogeneous biasing fields. Fourthly, the validation of the proposed approach was made by checking its accuracy and convergence through numerical examples. Besides, we proved analytically and numerically that it is the pressure difference, not the internal or external pressure separately, that determines the dispersion relations. Finally, we provided a detailed numerical evaluation to elucidate the static nonlinear response and the effects of the material gradient, pressure difference, and axial pre-stretch on the T and L wave propagation characteristics.

For the axisymmetric deformation response of the FG-EHC to the combined action of the pre-stretch and the pressure difference, we demonstrated that: 1) a uniform deformation state always exists even if the pre-stretched EHC has an FG behavior when there is no pressure; 2) to ensure the axisymmetric deformation and avoid the instability or bifurcation, there exists an allowable range of the pressure difference depending on the axial pre-stretch and the material gradient, which may be enlarged by increasing the material gradient; 3) the increasing material gradient stiffens the FG-EHC.

For the T waves propagating in the pressurized FG-EHC, we showed that: 1) the frequency increases significantly with the pre-stretch for the homogeneous EHC without the pressure difference, and the fundamental mode for a non-zero pressure difference always passes through the origin and is non-dispersive for the homogeneous EHC in the entire wave number range; 2) the fundamental mode for an FG-EHC is dispersive, even when the pressure difference vanishes, and the increasing pressure difference raises the curve slope or the group velocity of the fundamental mode; 3) the frequencies for all modes increase monotonically with the applied pressure difference and the material gradient; 4) the frequency range tuned by the pressure difference is enhanced and the gaps between two neighboring modes become larger when the material gradient increases; 5) the displacements in the long wavelength limit show a linear variation along the radial coordinate and the mode shape for short waves may be used to characterize the material property and the stress/strain state.

For the L waves, we can draw the following conclusions: 1) analogous to the T waves, the frequency and the phase velocity increase dramatically with their gaps between two neighboring modes becoming larger when the material gradient increases, and the material gradient raises noticeably the group velocities; 2) all modes are dispersive in all cases, even for a homogeneous EHC, and the cutoff frequency depends on the material gradient and the pressure difference; 3) the phase velocity of the fundamental mode in all cases asymptotically approachs the modified Rayleigh surface wave velocity at a large wave number and exhibits a non-monotonic behavior as a result of the complex wave interaction with the geometric boundaries; 4) the frequencies for all modes increase dramatically and monotonically with the material gradient for a positive pressure difference, whereas most modal frequencies for a negative pressure difference exhibit a non-monotonic variation with the material gradient which is a consequence of the competition between the changes in the material gradient and the amount of the nonlinear deformation; 5) similar to the T waves, the tunable frequency range by the pressure difference becomes larger when the material gradient increases; 6) the complex effect of the pressure difference on the L wave propagation characteristics depends strongly on the wave mode, the wave number range, the material gradient, and the axial pre-stretch; 7) the lowest L wave mode in the limit of the long waves corresponds to the uniform extensional mode, while that in the short wavelength limit behaves like Rayleigh-type surface waves either on the outer or inner surface.

All these results lay a theoretical foundation for the in-situ ultrasonic nondestructive evaluation to characterize the property, the pre-stress state, and the structural defects such as cracks of the FG-EHC and for the material tailoring to realize tunable FG-EHC waveguides through adjusting the pre-stretch and the pressure difference. Further research works, especially experiments, are required concerning these two interesting topics. Another interesting topic is to manipulate elastic waves in soft electroactive materials and structures by the material inhomogeneity and large deformation along with the electromechanical coupling, which should have wide applications in smart devices and intelligent soft robotics [66,67].

**Acknowledgments**

The work was supported by the National Natural Science Foundation of China (Nos. 11532001, 11621062, and 11402310), the German Research Foundation (DFG, Project-No: ZH 15/20-1), and the China Scholarship Council (CSC). Partial support from the Fundamental Research Funds for the Central Universities (No. 2016XZZX001-05) is also acknowledged.

## Appendix A. Elements of the system matrix $\mathbf{M}$

The elements of the system matrix $\mathbf{M}$ in the state equation (36) are given by

$$\mathbf{M}_{11} = \begin{bmatrix} -\dfrac{1}{r} & -\dfrac{1}{r}\dfrac{\partial}{\partial\theta} & -\dfrac{\partial}{\partial z} \\[2ex] -\dfrac{f_1}{r}\dfrac{\partial}{\partial\theta} & \dfrac{f_1}{r} & 0 \\[2ex] -f_2\dfrac{\partial}{\partial z} & 0 & 0 \end{bmatrix}, \quad \mathbf{M}_{12} = \begin{bmatrix} 0 & 0 & 0 \\[2ex] 0 & \dfrac{1}{\Gamma_{01212}} & 0 \\[2ex] 0 & 0 & \dfrac{1}{\Gamma_{01313}} \end{bmatrix}, \quad \mathbf{M}_{22} = \begin{bmatrix} 0 & -\dfrac{f_1}{r}\dfrac{\partial}{\partial\theta} & -f_2\dfrac{\partial}{\partial z} \\[2ex] -\dfrac{1}{r}\dfrac{\partial}{\partial\theta} & -\dfrac{f_1+1}{r} & 0 \\[2ex] -\dfrac{\partial}{\partial z} & 0 & -\dfrac{1}{r} \end{bmatrix},$$

$$\mathbf{M}_{21} = \begin{bmatrix} \rho\dfrac{\partial^2}{\partial t^2} - \dfrac{f_6}{r^2}\dfrac{\partial^2}{\partial\theta^2} + \dfrac{f_3}{r^2} - f_7\dfrac{\partial^2}{\partial z^2} & \dfrac{f_3+f_6}{r^2}\dfrac{\partial}{\partial\theta} & \dfrac{f_4}{r}\dfrac{\partial}{\partial z} \\[3ex] -\dfrac{f_3+f_6}{r^2}\dfrac{\partial}{\partial\theta} & \rho\dfrac{\partial^2}{\partial t^2} - \dfrac{f_3}{r^2}\dfrac{\partial^2}{\partial\theta^2} + \dfrac{f_6}{r^2} - \Gamma_{03232}\dfrac{\partial^2}{\partial z^2} & -\dfrac{f_8}{r}\dfrac{\partial^2}{\partial\theta\partial z} \\[3ex] -\dfrac{f_4}{r}\dfrac{\partial}{\partial z} & -\dfrac{f_8}{r}\dfrac{\partial^2}{\partial\theta\partial z} & \rho\dfrac{\partial^2}{\partial t^2} - f_5\dfrac{\partial^2}{\partial z^2} - \dfrac{\Gamma_{02323}}{r^2}\dfrac{\partial^2}{\partial\theta^2} \end{bmatrix}$$

where

$$f_1 = (\Gamma_{01221} + q)/\Gamma_{01212}, \quad f_2 = (\Gamma_{01331} + q)/\Gamma_{01313}, \quad f_3 = \Gamma_{01111} + \Gamma_{02222} - 2\Gamma_{01122} + 2q,$$
$$f_4 = \Gamma_{01111} + \Gamma_{02233} - \Gamma_{01122} - \Gamma_{01133} + q, \quad f_5 = \Gamma_{01111} + \Gamma_{03333} - 2\Gamma_{01133} + 2q,$$
$$f_6 = \Gamma_{02121} - (\Gamma_{01221} + q)^2/\Gamma_{01212}, \quad f_7 = \Gamma_{03131} - (\Gamma_{01331} + q)^2/\Gamma_{01313}, \quad f_8 = f_4 + \Gamma_{02332} + q$$

## Appendix B. Proof: Dispersion relations depend on the pressure difference $\Delta p$ only.

In this appendix, we prove that it is the pressure difference, not the internal or external pressure separately, that determines the dispersion relations of the incremental non-axisymmetric waves propagating in a pressurized FG-EHC. Instead of the SSM, we will adopt the conventional displacement method here.

The incremental equations and the associated boundary conditions were given in Eqs. (31)-(34) and Eq. (36). The traveling wave solution can be assumed in the following form

$$\left[ u_r, u_\theta, u_z, \dot{q} \right] = \left[ u(r), v(r), w(r), \dot{Q}(r) \right] \mathrm{e}^{\mathrm{i}(m\theta + kz - \omega t)} \tag{B.1}$$

which represents the incremental *non-axisymmetric* waves and can be reduced to the T and L waves by taking $m = 0$. Substituting Eq. (B.1) into Eqs. (31)-(32) and Eq. (34) yields

$$u' = -(\mathrm{i}mv + u)/r - \mathrm{i}kw, \quad u'' = -(\mathrm{i}mv' + u')/r + (\mathrm{i}mv + u)/r^2 - \mathrm{i}kw', \tag{B.2}$$

and

$$\dot{T}_{0rr} = \left[ (\Gamma_{01111} + q)u' + \Gamma_{01122}(imv + u)/r + \Gamma_{01133}ikw - \dot{Q} \right] \mathrm{e}^{\mathrm{i}(m\theta + kz - \omega t)},$$

$$\dot{T}_{0\theta\theta} = \left[ \Gamma_{01122}u' + (\Gamma_{02222} + q)(imv + u)/r + \Gamma_{02233}ikw - \dot{Q} \right] \mathrm{e}^{\mathrm{i}(m\theta + kz - \omega t)},$$

$$\dot{T}_{0zz} = \left[ \Gamma_{01133}u' + \Gamma_{02233}(imv + u)/r + (\Gamma_{03333} + q)ikw - \dot{Q} \right] \mathrm{e}^{\mathrm{i}(m\theta + kz - \omega t)},$$

$$\dot{T}_{0r\theta} = \left[ \Gamma_{01212}v' + (\Gamma_{01221} + q)(imu - v)/r \right] \mathrm{e}^{\mathrm{i}(m\theta + kz - \omega t)},$$

$$\dot{T}_{0\theta r} = \left[ \Gamma_{02121}(imu - v)/r + (\Gamma_{01221} + q)v' \right] \mathrm{e}^{\mathrm{i}(m\theta + kz - \omega t)}, \tag{B.3}$$

$$\dot{T}_{0rz} = \left[ (\Gamma_{01331} + q)iku + \Gamma_{01313}w' \right] \mathrm{e}^{\mathrm{i}(m\theta + kz - \omega t)},$$

$$\dot{T}_{0zr} = \left[ \Gamma_{03131}iku + (\Gamma_{01331} + q)w' \right] \mathrm{e}^{\mathrm{i}(m\theta + kz - \omega t)},$$

$$\dot{T}_{0\theta z} = \left[ \Gamma_{02323}imw/r + (\Gamma_{02332} + q)ikv \right] \mathrm{e}^{\mathrm{i}(m\theta + kz - \omega t)},$$

$$\dot{T}_{0z\theta} = \left[ \Gamma_{03232}ikv + (\Gamma_{02332} + q)imw/r \right] \mathrm{e}^{\mathrm{i}(m\theta + kz - \omega t)}$$

It should be pointed out that, according to Eq. (18), after the geometric parameter $\eta$, the material gradient parameter $\beta$, and the axial stretch $\lambda_z$ are given, the inner radius $a$ (or $\lambda_a$) can be determined analytically or numerically in terms of the pressure difference $\Delta p$. It means that the pressure difference completely determines the deformation state of the pressurized FG-EHC. Therefore, the instantaneous elasticity tensor in terms of the principal stretches is determined by the pressure difference from Eq. (8). On the other hand, the Lagrange multiplier $q$ is associated with the internal or external pressure from Eq. (4). However, due to the *constant* internal or external pressure, after inserting Eq. (B.3) into the incremental equations of motion (33), neglecting the derivatives of $q$ with respect to the radial coordinate $r$, and picking up those terms related to $q$, we can obtain from Eq. (33) the following terms:

$$\delta_1 = q\left\{ u'' + imv'/r + \left[ u' - (imv + u)/r \right]/r + ikw' \right\},$$

$$\delta_2 = q\left[ (imu' - v')/r - (imu - v)/r^2 + im(imv + u)/r^2 + v'/r + (imu - v)/r^2 + ikimw/r \right], \tag{B.4}$$

$$\delta_3 = q\left( iku' + ikvim/r + ikwik + iku/r \right)$$

which are all equal to zero due to Eq. (B.2). Therefore, for a general form of the strain energy function, the Lagrange multiplier $q$ does not appear in the incremental equations governing the incremental non-axisymmetric waves. Additionally, substituting Eq. (B.3) into the incremental boundary conditions (36), we have

$$(\Gamma_{01111} + q - p_a)u' + \Gamma_{01122}(imv + u)/r + \Gamma_{01133}ikw - \dot{Q} = 0,$$

$$(\Gamma_{01221} + q - p_a)(imu - v)/r - \Gamma_{01212}v' = 0, \quad (\Gamma_{01331} + q - p_a)iku + \Gamma_{01313}w' = 0 \tag{B.5}$$

where $p_a$ is the applied pressure on the inner or outer surface of the hollow cylinder (i.e., the internal or external pressure). Using Eq. (4) and the boundary conditions (17), the values of the Lagrange multiplier $q$ on the inner and outer surfaces can be calculated as

$$q(a) = (\lambda_1 W_1)\big|_{r=a} + p_{in}, \quad q(b) = (\lambda_1 W_1)\big|_{r=b} + p_{ou} \tag{B.6}$$

which indicates that the difference between the internal or external pressure and the Lagrange multiplier $q(a)$ or $q(b)$ on the inner or outer surface can be expressed by $(\lambda_1 W_1)\big|_{r=a}$ or $(\lambda_1 W_1)\big|_{r=b}$. It is therefore clear from Eq. (B.5) that the internal or external pressure will not appear separately in the boundary conditions.

In conclusion, for a general form of the strain energy function, only the pressure difference, but not the internal or external pressure separately, determines the dispersion relations of the incremental non-axisymmetric waves propagating in a pressurized FG-EHC.

# Figure Captions

**Fig. 1.** Schematic diagram of an FG-EHC with cylindrical coordinates and its cross sections: (a) undeformed configuration; (b) deformed configuration induced by a pressure difference and an axial pre-stretch.

**Fig. 2.** Variations of $\lambda_a$ with the dimensionless pressure difference $\Delta p^* = \Delta p / \mu_{10}$ in an FG-EHC for different values of the material gradient parameter $\beta$ and three different axial pre-stretches $\lambda_z$: (a) $\lambda_z = 0.8$; (b) $\lambda_z = 1$; (c) $\lambda_z = 2$.

**Fig. 3.** Comparison of the frequency spectra of the first five wave modes obtained by the exact solutions [62] and the SSM for a homogeneous EHC without internal or external pressure at different pre-stretches: (a) T waves; (b) L waves.

**Fig. 4.** Variations of the dimensionless cut-off frequency of the first three wave modes with the axial pre-stretch $\lambda_z$ for a homogeneous EHC without pressures: (a) T waves; (b) L waves.

**Fig. 5.** Variations of the second dimensionless frequency of the T waves (a, b) and the first dimensionless frequency of the L waves (c, d) at a wave number $K = 2.6$ with the number of the discretized thin layers for a pressurized FG-EHC with $\lambda_z = 1$ and two material gradient parameters $\beta = 5$ (a, c) and $\beta = 10$ (b, d) at different pressure differences.

**Fig. 6.** The frequency spectra of the first five wave modes obtained by the SSM for a pressurized FG-EHC with $\beta = 5$ and $\lambda_z = 1$ at a fixed pressure difference but at different internal pressures: (a) T waves ($\Delta p^* = 0$); (b) L waves ($\Delta p^* = -1$).

**Fig. 7.** The frequency spectra (each with the first five branches) of the T waves in a pressurized FG-EHC with $\lambda_z = 1$ at different pressure differences: (a) $\beta = 0$; (b) $\beta = 5$; (c) $\beta = 10$.

**Fig. 8.** Variations of the dimensionless frequency of the T wave modes with the pressure difference $\Delta p^*$ for a pressurized FG-EHC with $\lambda_z = 1$ and $\beta = 0$ (a, c) or $\beta = 5$ (b, d): (a, b) the first three nonzero cut-off freqeuncies; (c, d) the first four frequencies at a wave number $K = 5$.

**Fig. 9.** Normalized circumferential displacement variations of the lowest T wave mode with $\lambda_z = 1$ for different values of $\Delta p^*$ at two wave numbers $K = 0.2$ and $K = 10$: (a) $\beta = 0$; (b) $\beta = 5$.

**Fig. 10.** The frequency spectra of the T waves in a pressurized FG-EHC with $\lambda_z = 1$ at different material gradient parameters: (a) $\Delta p^* = -2$; (b) $\Delta p^* = 0$; (c) $\Delta p^* = 2$.

**Fig. 11.** Variations of the dimensionless frequency of the T wave modes with the material gradient parameter $\beta$ for a pressurized FG-EHC at different pressure differences and $\lambda_z = 1$ (a, c)

or $\lambda_z = 2$ (b, d): (a, b) the first four nonzero cut-off frequencies; (c, d) the first five frequencies at a wave number $K = 5$.

**Fig. 12.** Normalized circumferential displacement variations of the lowest T wave mode with $\lambda_z = 1$ for different values of $\beta$ at two wave numbers $K = 0.2$ and $K = 10$: (a) $\Delta p^* = 0$; (b) $\Delta p^* = -2$.

**Fig. 13.** The frequency spectra (a-c) and phase velocity spectra (d-f) of the L wave modes for a pressurized FG-EHC with $\lambda_z = 1$ at different material gradient parameters: (a, d) $\Delta p^* = -2$; (b, e) $\Delta p^* = 0$; (c, f) $\Delta p^* = 2$.

**Fig. 14.** Variations of the dimensionless frequency of the L wave modes with the material gradient parameter $\beta$ for a pressurized FG-EHC at different pressure differences and $\lambda_z = 1$ (a, c) or $\lambda_z = 2$ (b, d): (a, b) the first four nonzero cut-off frequencies; (c, d) the first five frequencies at a wave number $K = 5$.

**Fig. 15.** Normalized radial and axial displacement variations of the lowest L wave mode for $\lambda_z = 1$ and different values of $\beta$ at two wave numbers $K = 0.2$ (a, c) and $K = 10$ (b, d): (a, b) $\Delta p^* = 0$; (c, d) $\Delta p^* = -2$.

**Fig. 16.** The frequency spectra (a-c) and phase velocity spectra (d-f) of the first three L wave modes for a pressurized FG-EHC with $\lambda_z = 1$ at different pressure differences: (a, d) $\beta = 0$; (b, e) $\beta = 5$; (c, f) $\beta = 10$.

**Fig. 17.** Variations of the first two nonzero cut-off frequencies of the L wave modes with the pressure difference $\Delta p^*$ for a pressurized FG-EHC at $\lambda_z = 1$ and $\lambda_z = 2$ for two different material gradient parameters: (a) $\beta = 0$; (b) $\beta = 5$.

**Fig. 18.** Variations of the dimensionless frequency of the first three L wave modes at a small wave number $K = 2$ with the pressure difference $\Delta p^*$ for a pressurized FG-EHC at $\lambda_z = 1$ and $\lambda_z = 2$ for three different material gradient parameters: (a) $\beta = 0$; (b) $\beta = 2$; (c) $\beta = 5$.

**Fig. 19.** Variations of the dimensionless frequency of the first three L wave modes at a large wave number $K = 8$ with the pressure difference $\Delta p^*$ for a pressurized FG-EHC at $\lambda_z = 1$ (a-c) and $\lambda_z = 2$ (d-f) for different material gradient parameters: (a, d) $\beta = 0$; (b, e) $\beta = 2$; (c, f) $\beta = 5$.

**Fig. 20.** Normalized radial and axial displacement variations of the lowest L wave mode for $\lambda_z = 1$ and different values of $\Delta p^*$ at two wave numbers $K = 0.2$ (a, c) and $K = 10$ (b, d): (a, b) $\beta = 0$; (c, d) $\beta = 5$.

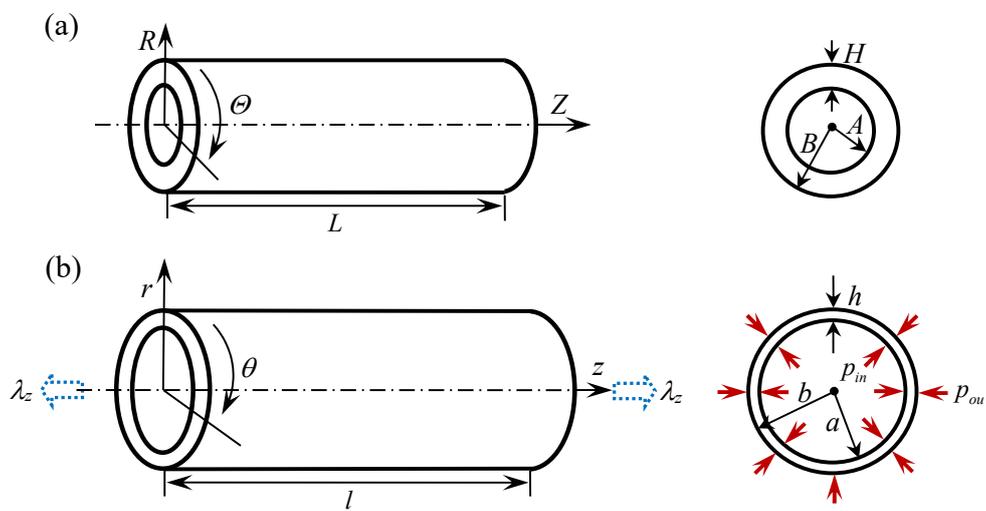

**Fig. 1**

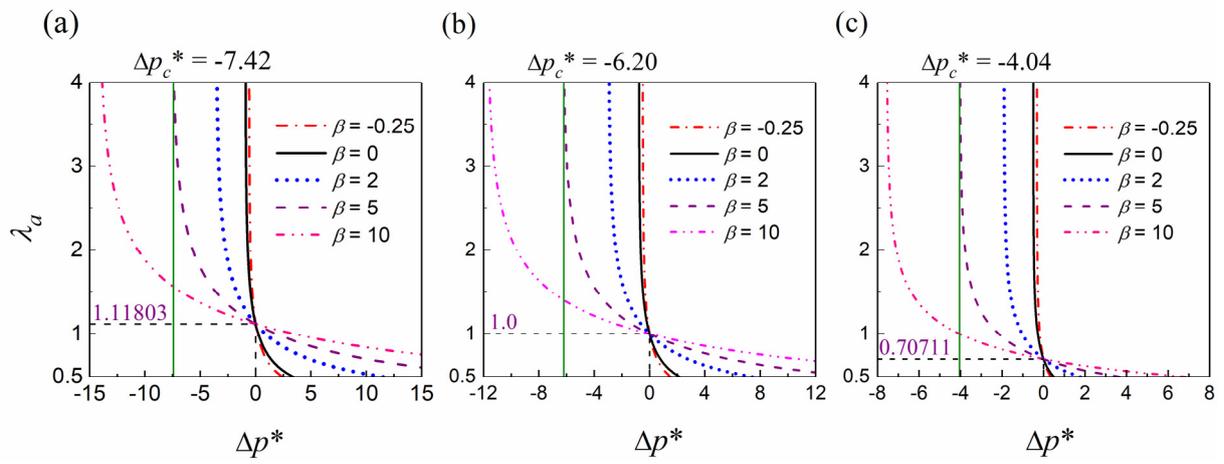

**Fig. 2**

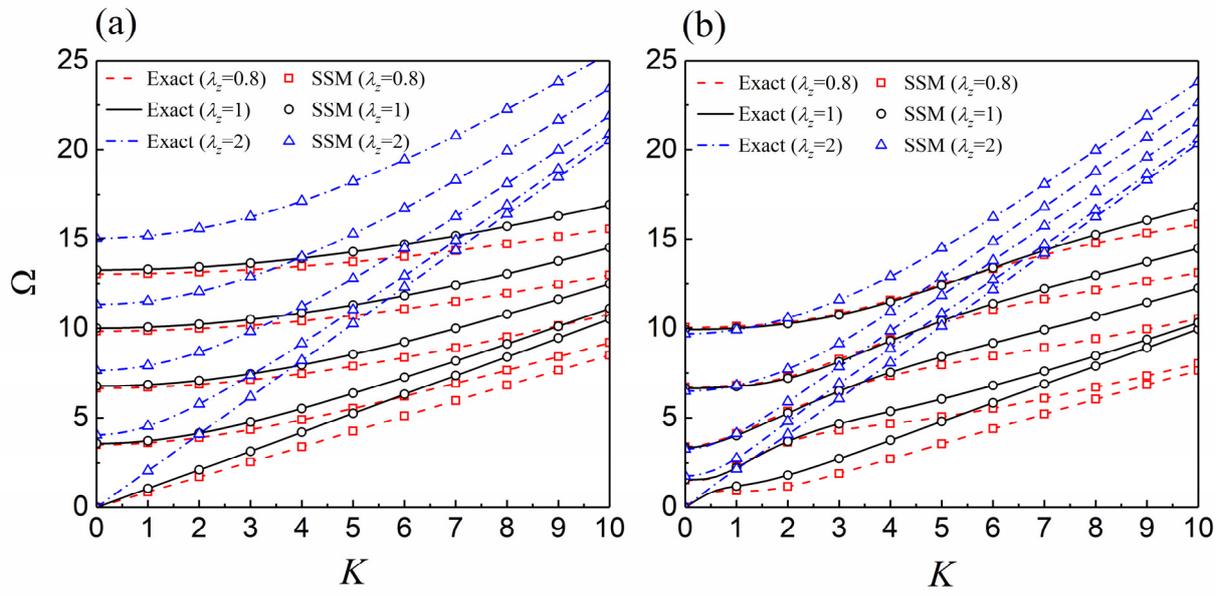

**Fig. 3**

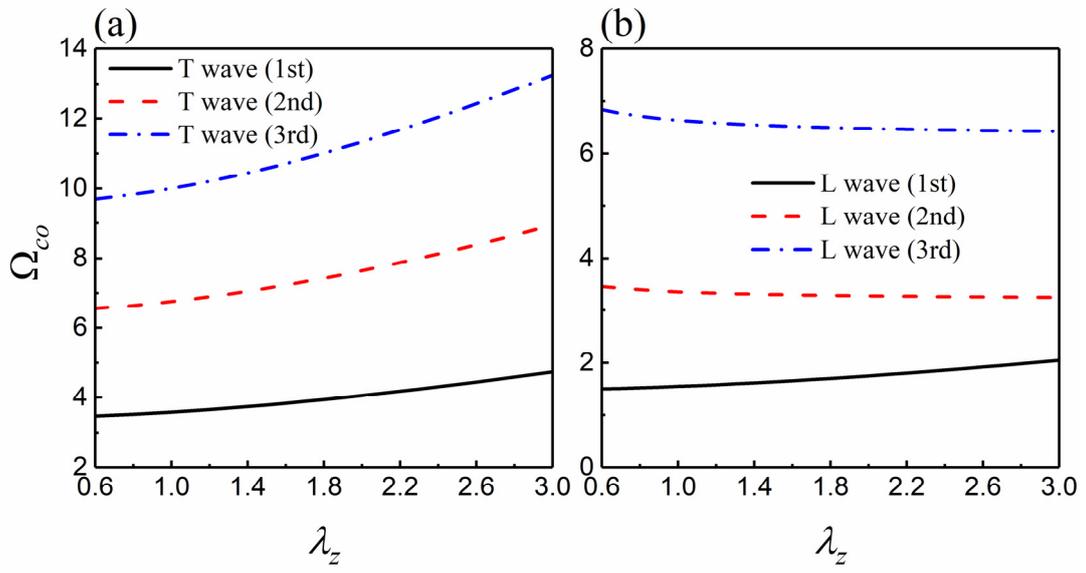

**Fig. 4**

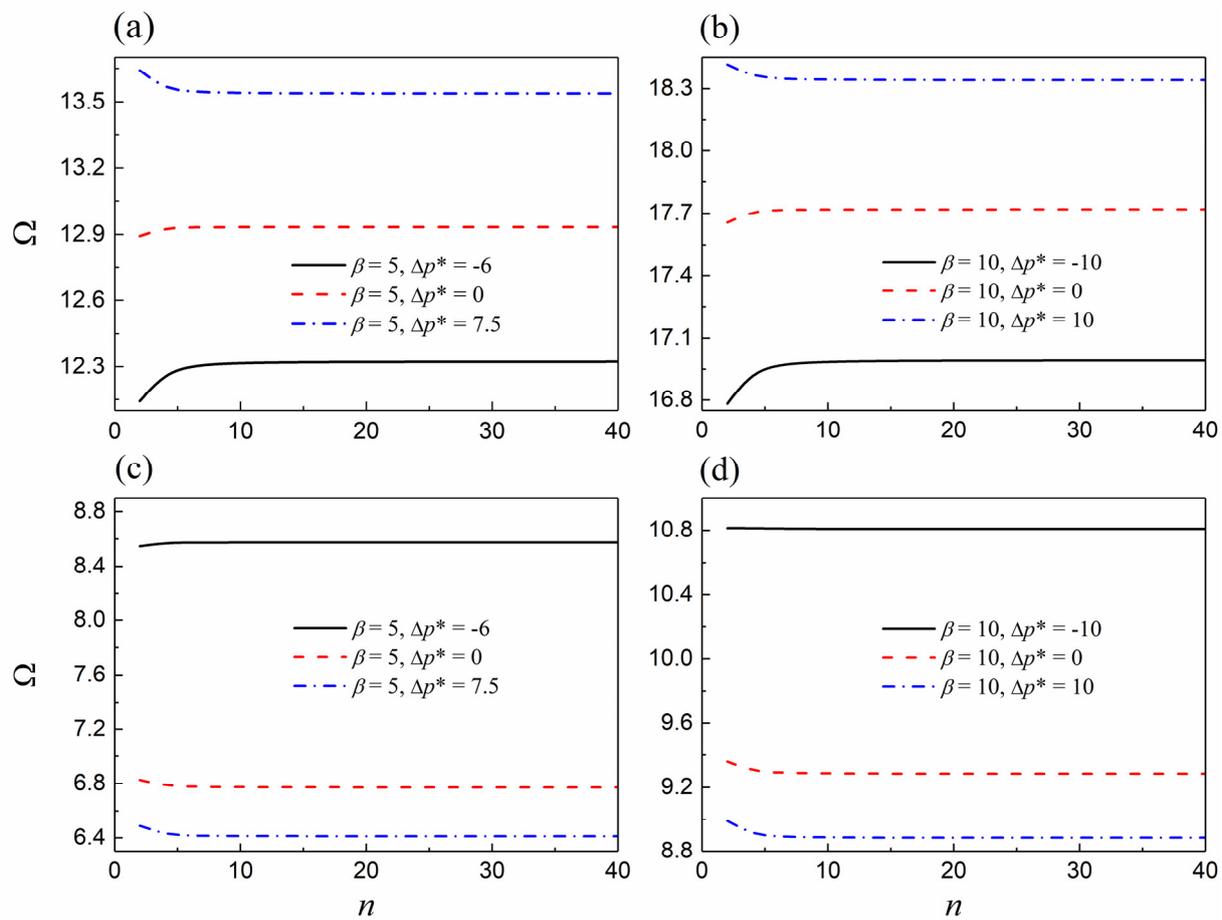

**Fig. 5**

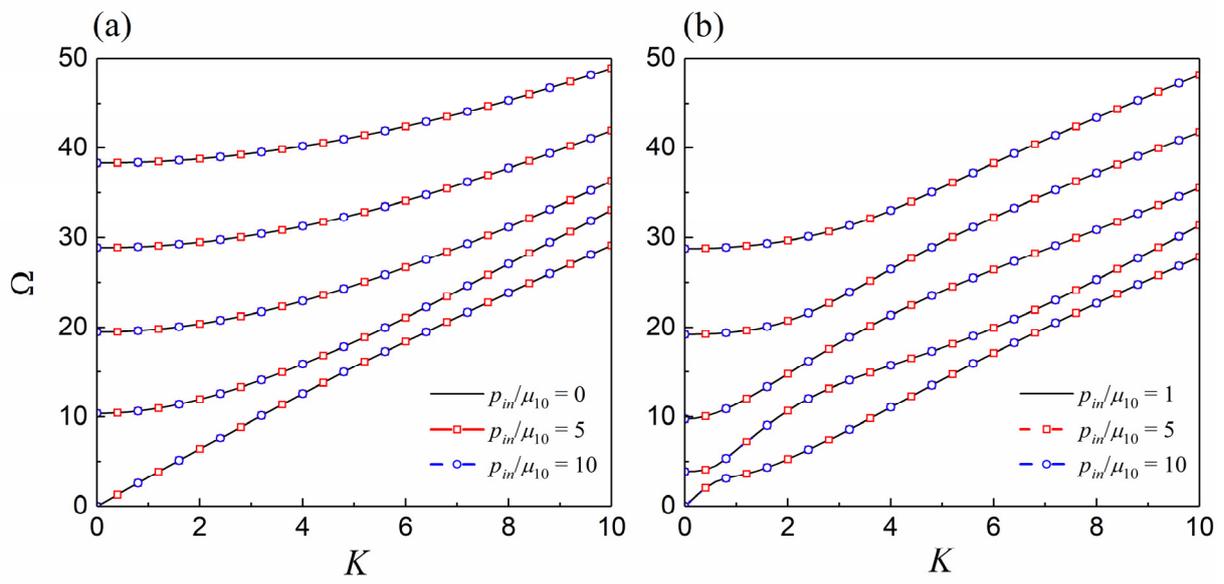

**Fig. 6**

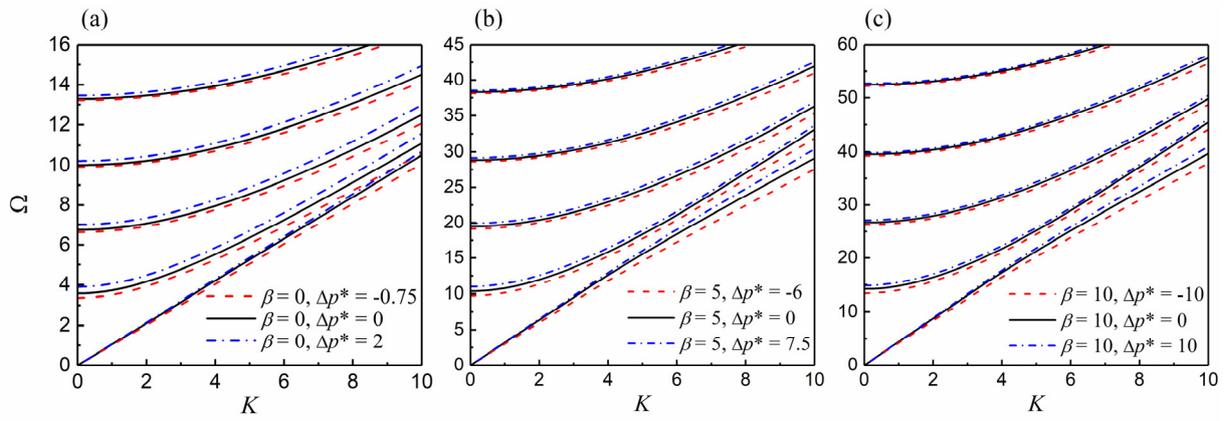

**Fig. 7**

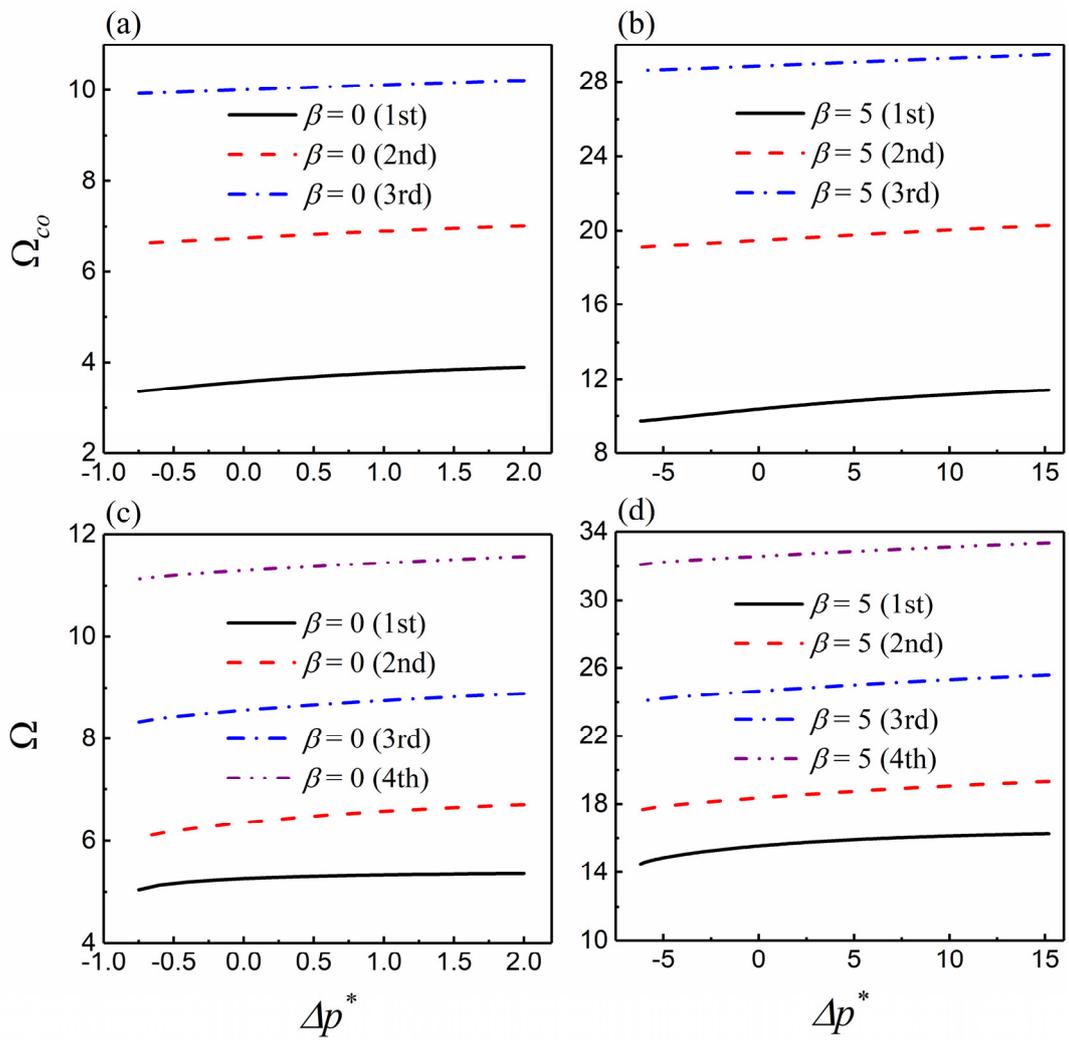

**Fig. 8**

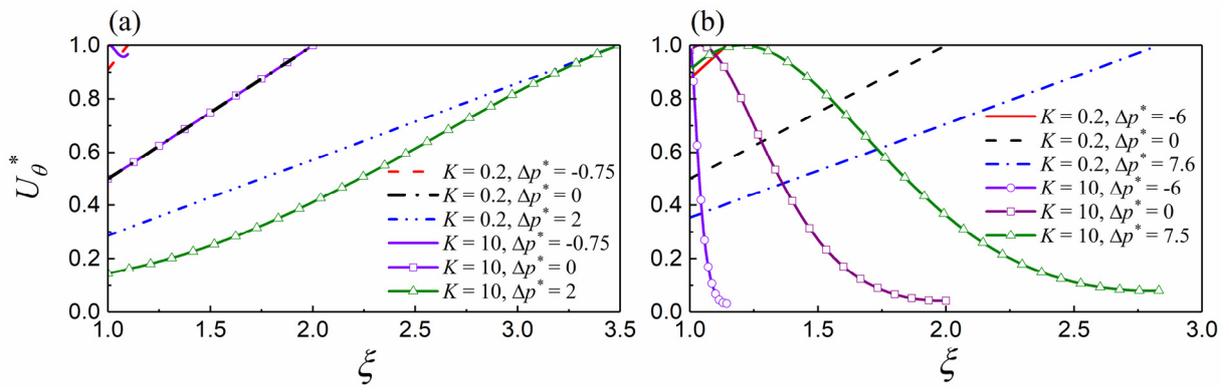

**Fig. 9**

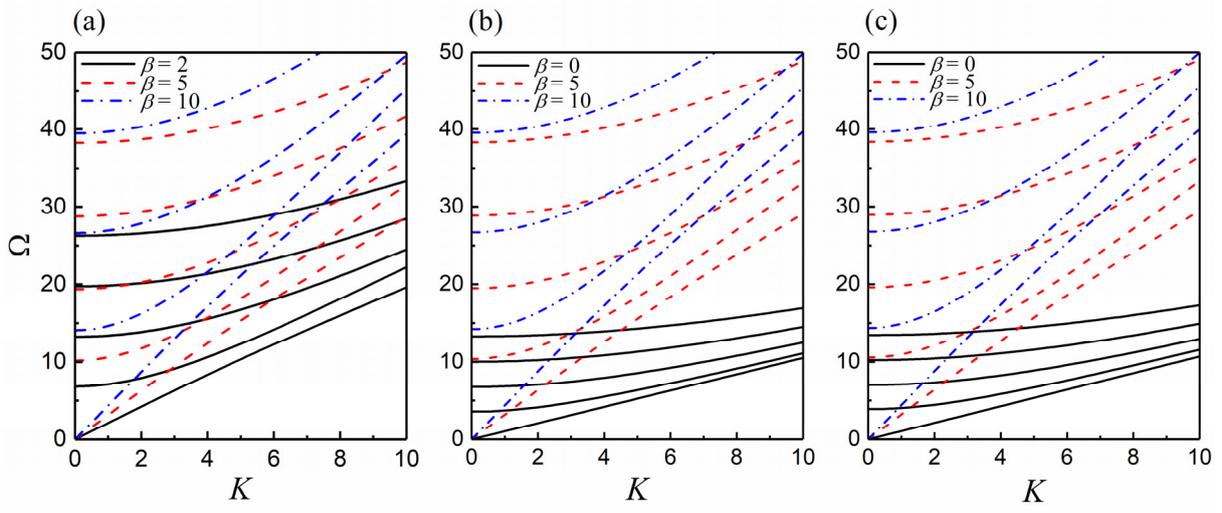

**Fig. 10**

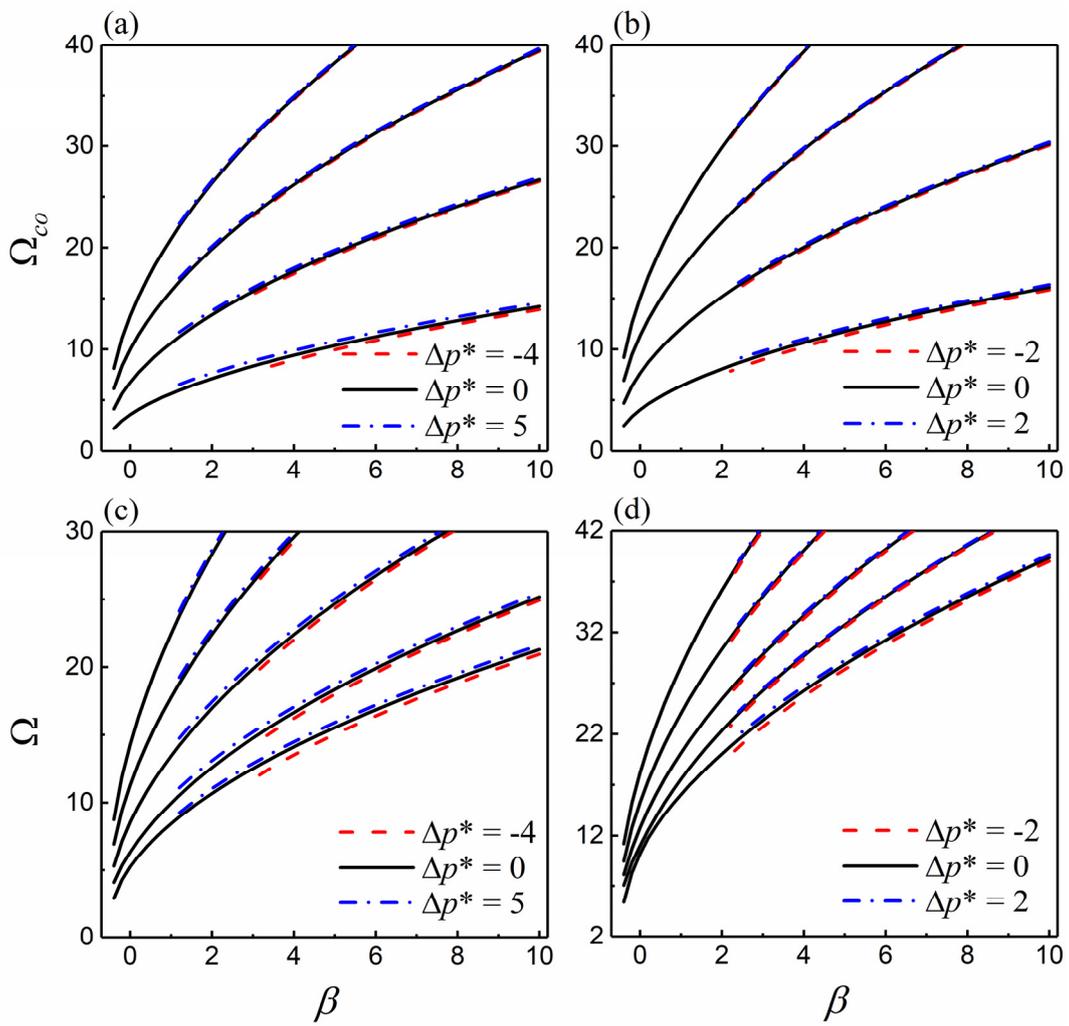

**Fig. 11**

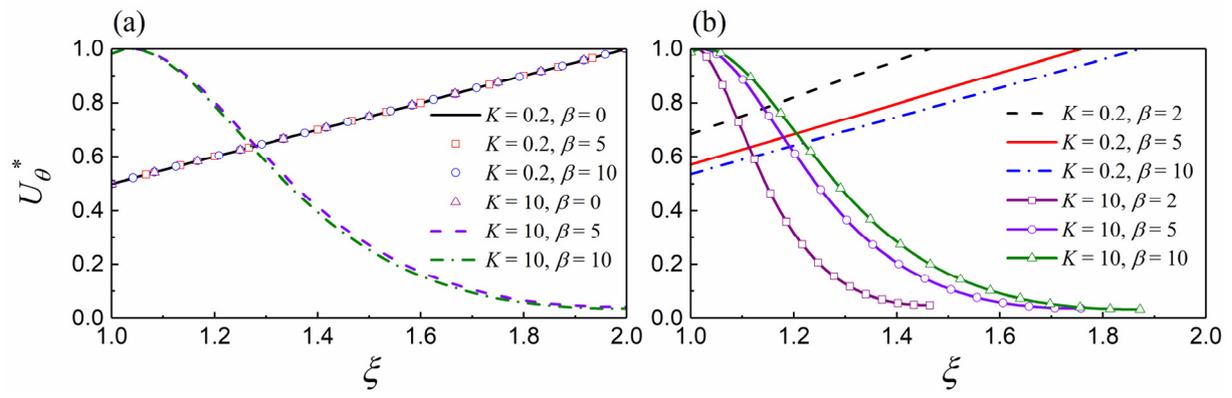

**Fig. 12**

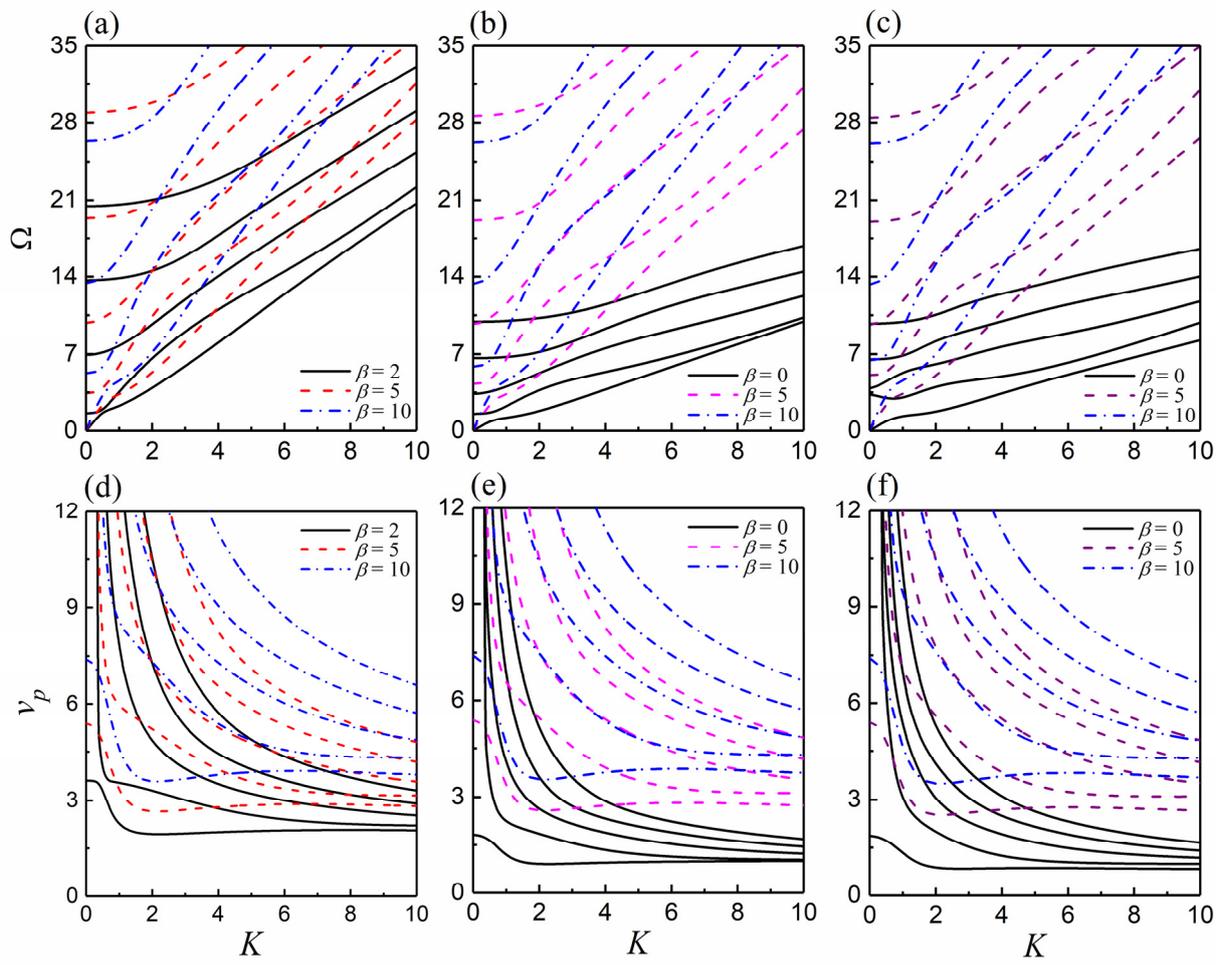

**Fig. 13**

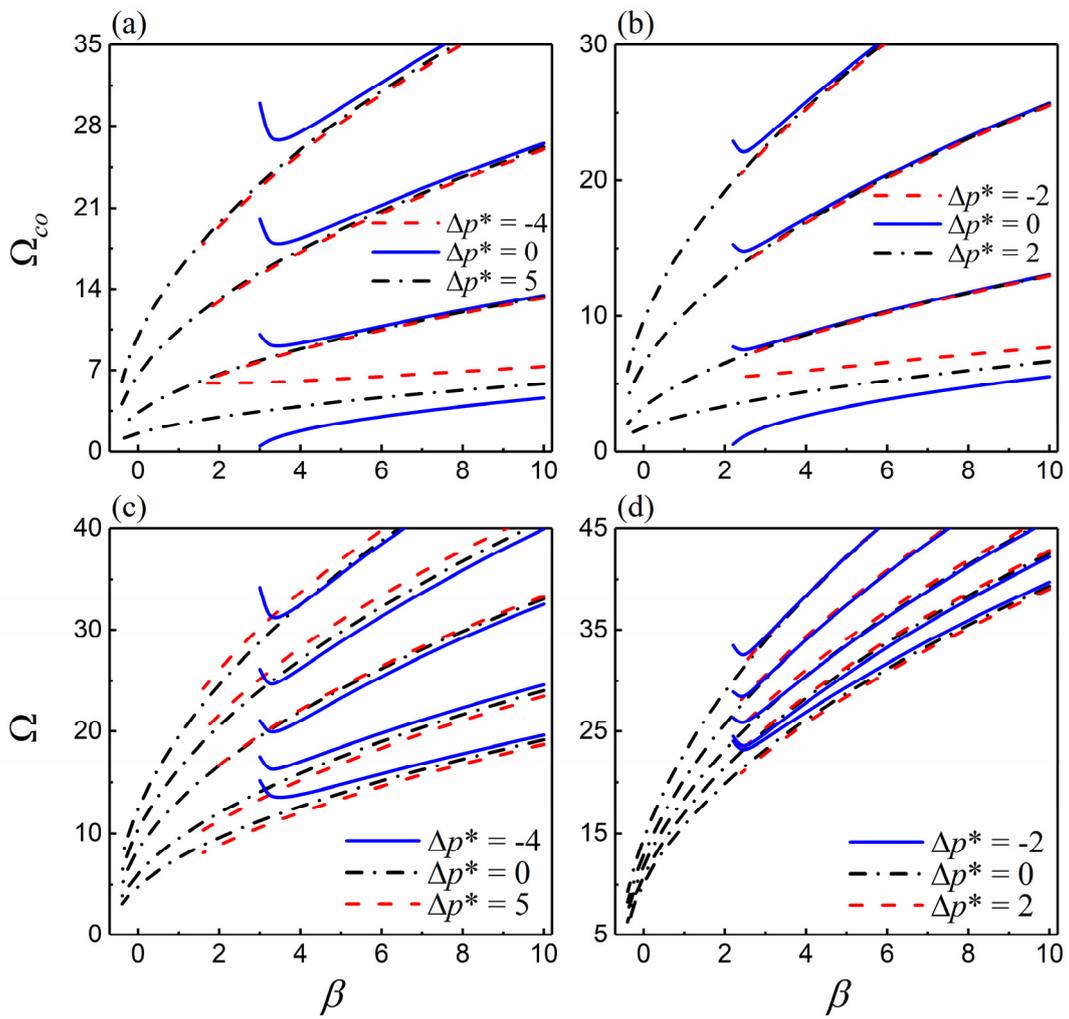

**Fig. 14**

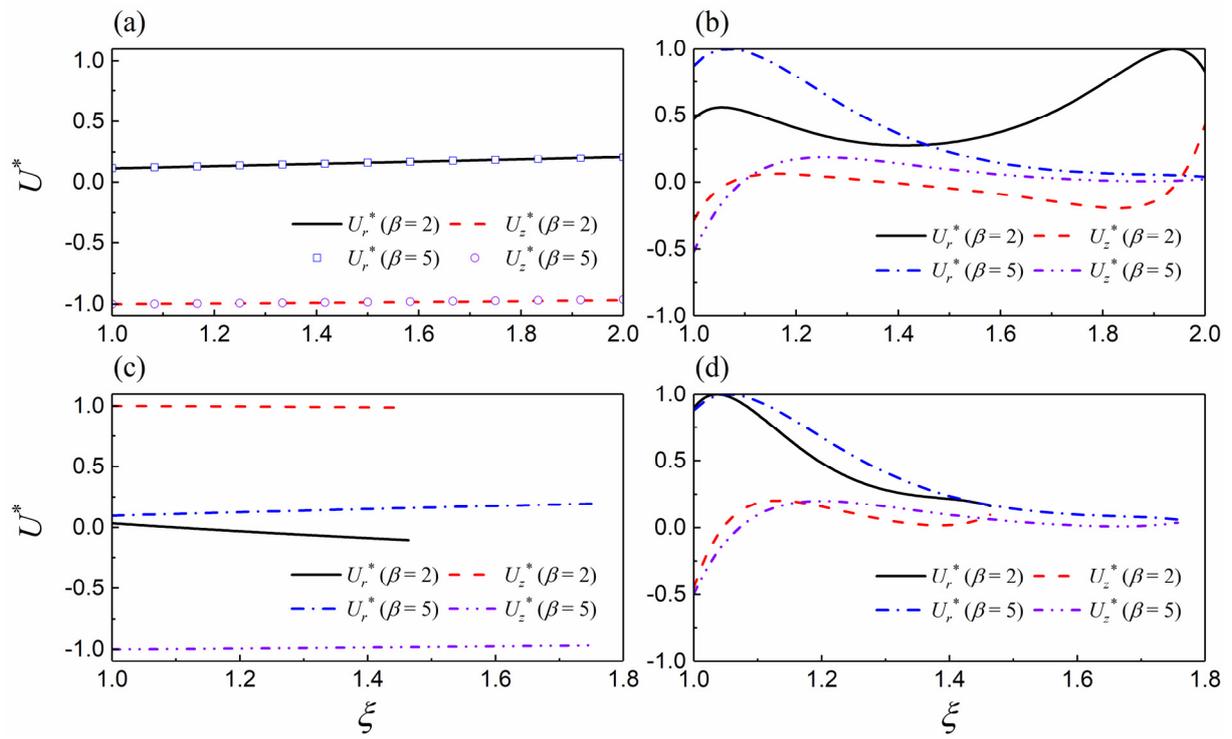

**Fig. 15**

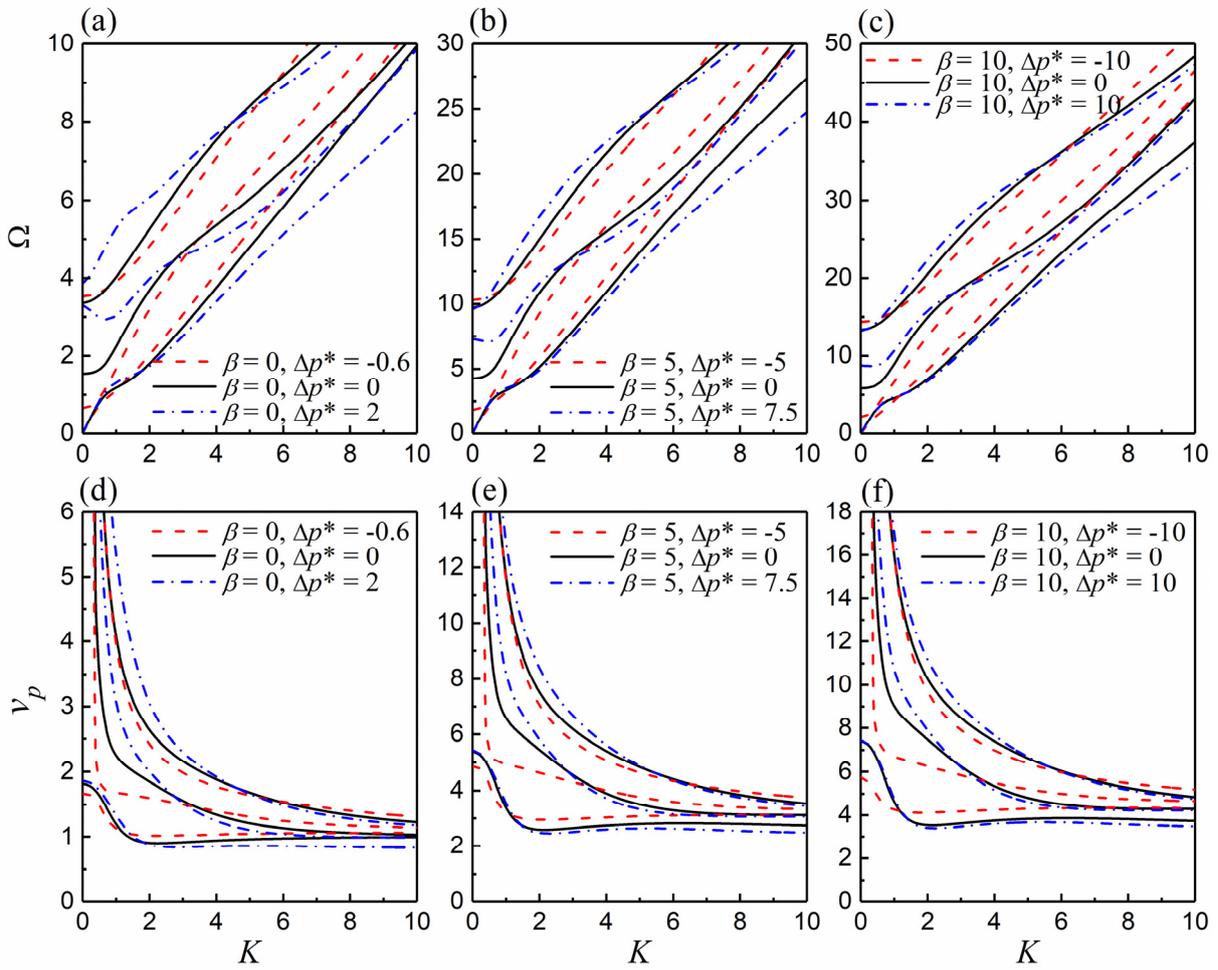

**Fig. 16**

为了整齐起见，请将图 c 中的标注放在右下角！放不下的话字体可以小一点！

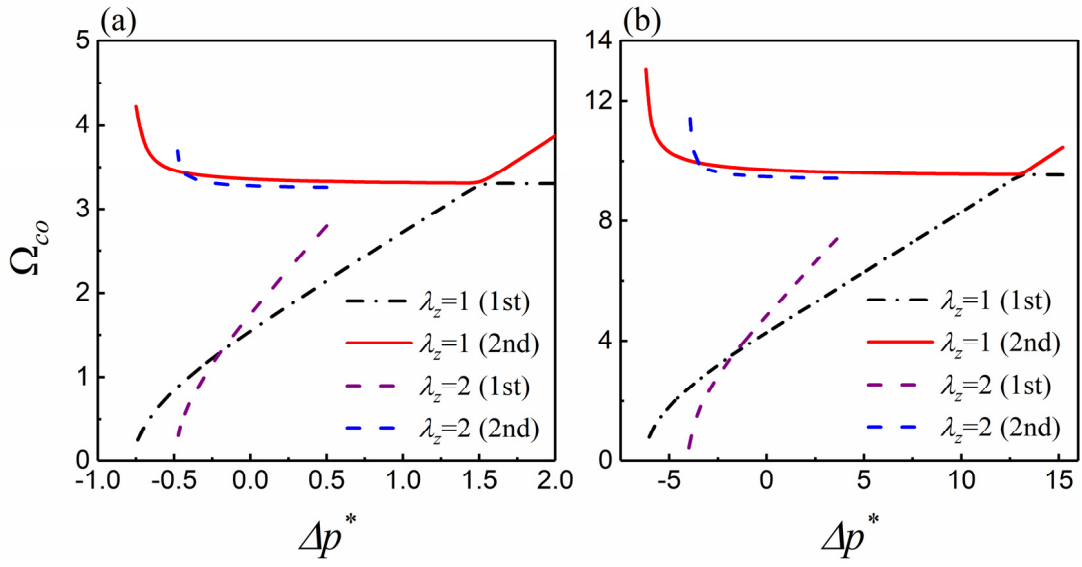

**Fig. 17**

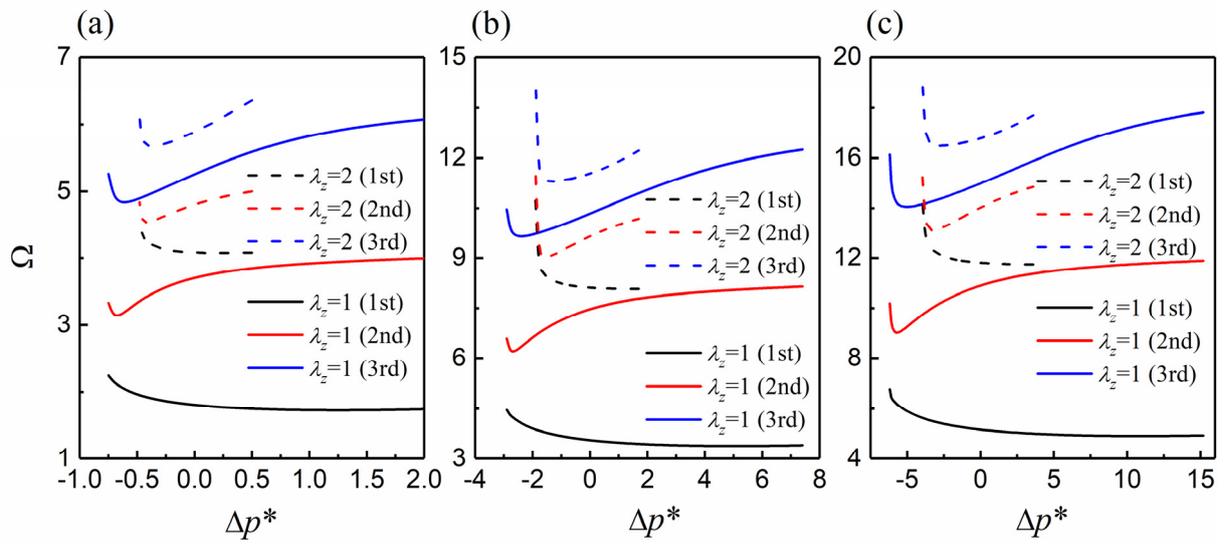

**Fig. 18**

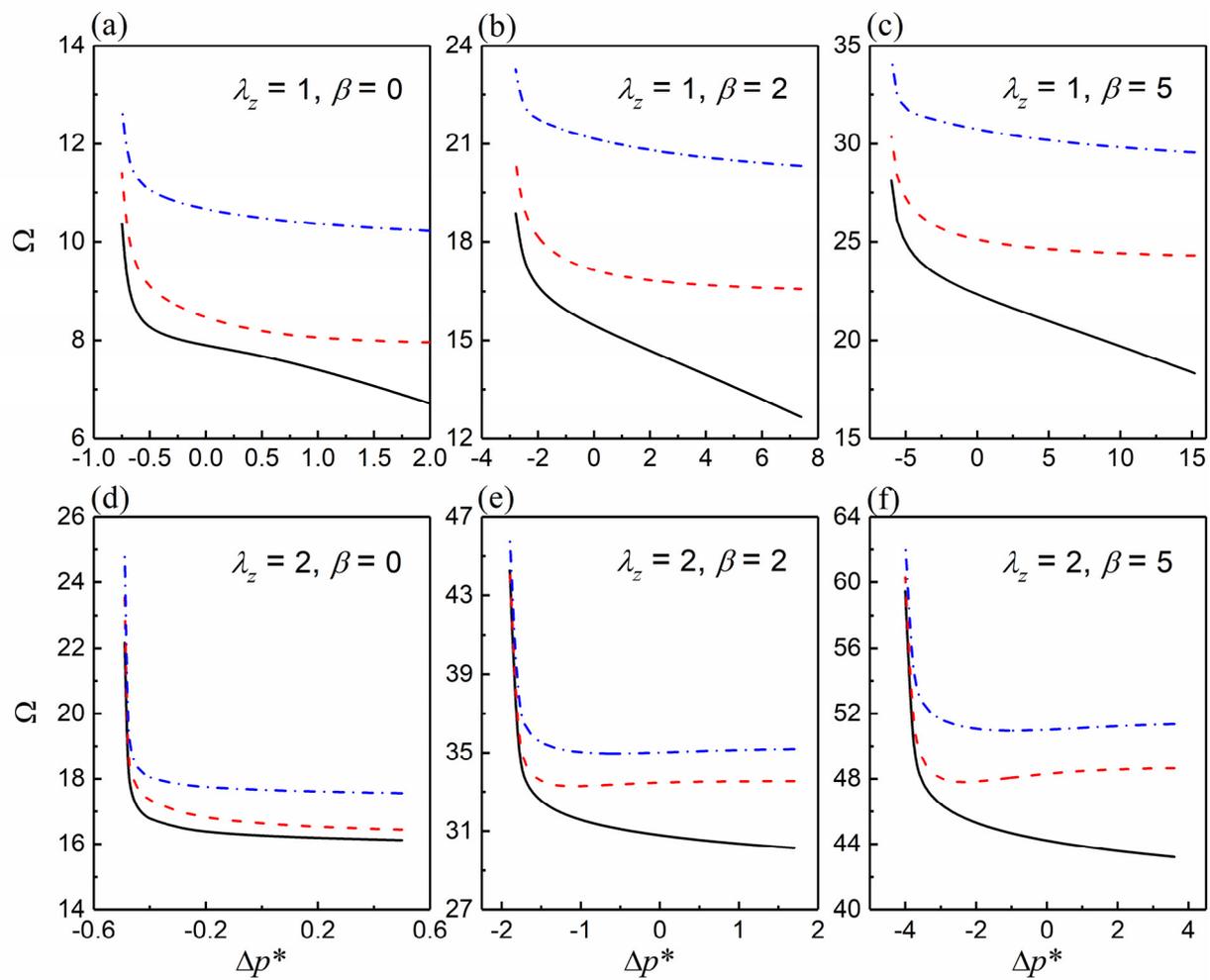

**Fig. 19**

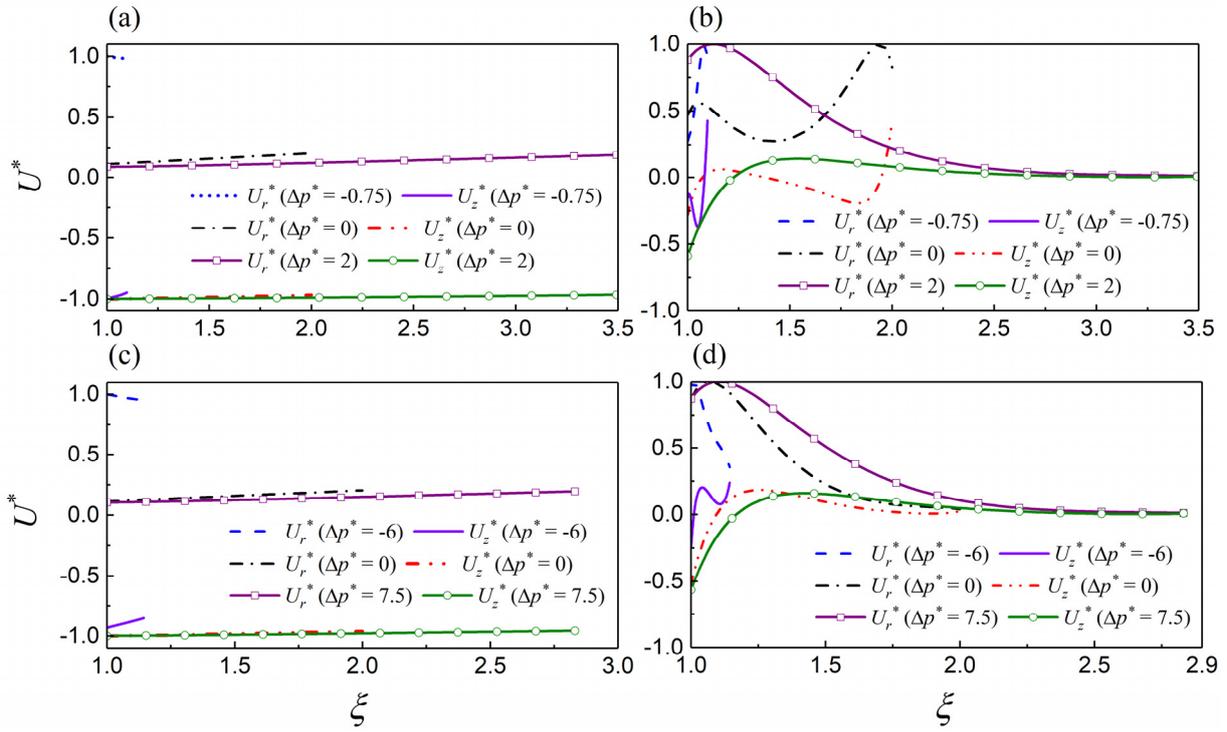

**Fig. 20**